\documentclass[structabstract]{aa}  
\usepackage{graphicx}
\usepackage{textgreek}
\usepackage{hyperref}
\usepackage{txfonts}
\begin{document}

   \title{Kinematic distinction of the two subpopulations of X-ray pulsars}

   \author{M. Pri\v{s}egen }

   \institute{Department of Theoretical Physics and Astrophysics, Masaryk University, Kotl\'{a}\v{r}sk\'{a} 2, 611 37
              Brno, Czech Republic\\
              \email{michalprisegen@gmail.com}
             }

   \date{Received ... / Accepted ...}

 
  \abstract
   {The population of Be/X-ray binaries shows strong evidence of bimodality, especially in the spin period of neutron stars. Several physical mechanisms may produce this bimodality. The most favored candidate mechanisms are two distinct supernova channels or different accretion modes of the neutron stars in Be/X-ray binaries. Investigating the kinematics of these systems may provide some additional insight into the physics of this bimodality.  }
   {If the two Be/X-ray binary subpopulations arise from two distinct supernova types, then the two subpopulations should have different peculiar (systemic) velocities. This can be tested either directly, by measuring the velocity of the system, or indirectly, by measuring the position of the system with respect to its birthplace. A difference in the peculiar velocity magnitude between the subpopulations would favor the supernova hypothesis, and the lack of this difference would suggest that the accretion hypothesis is a more favorable option to explain the bimodality.}
   {Using the most recent Gaia dataset and the newest catalogs of Small Magellanic Cloud (SMC) star clusters, we analyzed the tangential peculiar velocities of Be/X-ray binaries in the Galaxy and the positions of Be/X-ray binaries in the SMC. We used the distance of the system from the nearest young star cluster as a proxy to the tangential velocity of the system. We applied statistical testing to investigate whether the two subpopulations that are divided by the spin of the neutron star are also kinematically distinct.}
   {There is evidence that the two subpopulations are indeed kinematically distinct.  However, the tangential peculiar velocities of the two subpopulations are the reverse from what is expected from the distinct supernova channel hypothesis. We find some marginal evidence ($p \approx$~0.005) that the Galactic Be/X-ray binaries from the short-spin subpopulation have systematically higher peculiar velocities than the systems from the long-spin subpopulation. The same effect, but weaker, is also recovered for the SMC Be/X-ray binaries for all considered cluster catalogs. The unexpected difference in the peculiar velocities between the two subpopulations of Be/X-ray binaries contradicts these two hypotheses, and an alternative physical explanation for this may be needed.}
   {}


   \maketitle
%

\section{Introduction}
Be/X-ray binaries (BeXRBs) are the most numerous subclass of high-mass X-ray binaries in the Galaxy. They are systems containing a neutron star with a mass-losing Be-type main-sequence companion that is surrounded by a circumstellar decretion disk (e.g., Rivinius et al. \cite{rivinius_2013}). These objects are typically revealed by X-ray activity that is fueled by mass accretion. Most of the mass accretion takes place during periastron passages, when the neutron star passes in the vicinity, in some cases, even through, the decretion disk of the Be star (e.g., Ziolkowski \cite{ziolkowski_2002}; Reig \cite{reig_2011}; Casares \cite{casares_2017}).

A supernova explosion occurring in a massive binary leads to a disruption of the system in the majority of cases (see, e.g., Brandt \& Podsiadlowski \cite{brandt_95}; De Donder \cite{de_donder_97}; Eldridge et al. \cite{eldridge_2011}; Renzo et al. \cite{renzo_2019}). For the systems that remain bound, BeXRBs provide a valuable but inherently biased laboratory for studying the physics of the supernova explosions that formed their neutron stars. BeXRBs provide a well-defined and simple population: each hosts a neutron star primary with a mass of $\sim 1.4 \, \mathrm{M_{\odot}}$ (with the notable exception of MWC 656, which hosts a black hole; Casares et al. \cite{casares_2014}) and a secondary star from a relatively narrow spectral distribution that peaks at B0 (Reig \cite{reig_2011}). This population  nevertheless exhibits a wide variety of properties that encode the information about the past supernova event in the system and the massive binary progenitor. Moreover, the short lifetime of the BeXRB phase, typically $\sim 10$ Myr (van den Heuvel et al. \cite{heuvel_2000}), does not allow parameters such as the neutron star masses, orbital periods, and peculiar velocities to change significantly. These parameters are therefore close to their birth values just after the supernova explosion.

The original idea that there might be subpopulations in the BeXRB population was proposed by Pfahl et al. (\cite{pfahl_2002}), who reported a subclass of BeXRBs with low eccentricities and low X-ray luminosities. They proposed that this subclass originates from the binaries where the initially more massive star undergoing a supernova explosion has a rapidly rotating core, which results in a neutron star that has received only a small natal kick. Podsiadlowski et al. (\cite{podsiadlowski_2004}) and van den Heuvel (\cite{heuvel_2004}) proposed that this low-eccentricity subpopulation might be explained if the neutron stars in these systems underwent an electron-capture supernova (ECSN). The ECSNe are the result of the collapse of an oxygen-neon-magnesium core of a lower mass star (possibly with an initial mass as low as 6~$\mathrm{M_{\odot}}$ if it is in a tight binary, especially at lower metallicities; Podsiadlowski et al. \cite{podsiadlowski_2004}) as it loses pressure support owing to the sudden capture of electrons by neon or magnesium nuclei (Nomoto \cite{nomoto_1984}; Nomoto \cite{nomoto_1987}), ejecting little mass in the supernova explosion (normally $\lesssim 1 \mathrm{M_{\odot}}$) and imparting little to no kick to the nascent neutron star (Podsiadlowski et al. \cite{podsiadlowski_2004}; van den Heuvel \cite{heuvel_2004}). The classical high-eccentricity BeXRB population would then be a result of iron-core-collapse supernovae (CCSNe), which occur after a degenerate iron core forms inside a higher mass star (e.g., Cerda-Duran \& Elias-Rosa \cite{cerda-duran_2018} and the references therein). CCSNe eject more mass and might impart a substantial kick to the newly formed neutron stars as well. The evidence supporting the existence of two distinct explosion mechanisms is not limited to BeXRBs. Observations of double neutron stars, the bimodal velocity distribution of young pulsars, and a high number of neutron stars retained in globular clusters can be attributed to two supernova explosion mechanisms (Beniamini \& Piran \cite{beniamini_2016}; Verbunt et al. \cite{verbunt_2017}; Pfahl et al. \cite{pfahl_2002}). 

Examining the spin--orbital period diagram of BeXRBs, Knigge et al. (hereafter KCP; \cite{knigge_2011}) also noted that the BeXRB population consists of two subpopulations, a short-period subpopulation with a characteristic orbital period of $P_{\mathrm{orb}} \approx$~40~d and spin period $P_{\mathrm{s}} \approx$~10~s, and a long-period subpopulation with $P_{\mathrm{orb}} \approx$~100~d and $P_{\mathrm{s}} \approx$~200~s. The histogram of $P_{\mathrm{orb}}$ and $P_{\mathrm{s}}$ was used to estimate an approximate threshold dividing the subpopulations. This threshold lies at 60~d and 40~s, respectively. Especially in the case of spin periods, the dip in the histogram is considerably wide, therefore these values need to be considered with caution. The two subpopulations are more clearly separated in the spin period than in the orbital period. KCP theorized that these two subpopulations are also a consquence of two different supernovae types that occur in these binaries, where the ECSNe reportedly produce the short-period subpopulation and the CCSNe produce the long-period subpopulation. The observed BeXRB spin periods are not a direct result of the supernova explosion itself, rather, they evolve during the BeXRB phase toward some equilibrium spin period $\mathrm{P_{eq}}$. This $\mathrm{P_{eq}}$ is dependent on the orbital period of the BeXRBs, which produces the well-known correlation in the spin--orbital period diagram, but it also depends on other parameters that are expected to be different for ECSNe and CCSNe (e.g., resulting neutron star magnetic field and neutron star mass; Waters  \& van Kerkwijk \cite{waters_89}).

Cheng et al. (\cite{cheng_2014}) proposed an alternative explanation for the two subpopulations in $P_{\mathrm{s}}$, where the bimodality in $P_{\mathrm{s}}$ can be ascribed to different accretion modes of the neutron stars in BeXRBs. Here, the BeXRB systems that exhibit giant outbursts tend to have shorter spins. During giant outbursts, the neutron star accretes from a thin disk with a relatively long lifetime, which efficiently transfers mass and angular momentum to the neutron star, so that its spin period reaches $P_{\mathrm{s}}$~$\sim$~10~s. For the BeXRBs, which undergo predominantly normal outbursts or have no outbursts at all, the accretion torques are smaller. The accretion flows around the neutron stars within these systems are in the form of advection-dominated accretion flows, meaning that the spin-up is infrequent and ineffective. The sources from this subpopulation then exhibit spin periods of about $P_{\mathrm{s}}$~$\sim$~100~s. While the supernova mechanism has some effect on the occurrence and type of outbursts and thus spin periods, it is unclear how dominant it is and what other processes are relevant in this case.

These two competing hypotheses can be tested using the BeXRB parameters that do not depend, or depend only negligibly, on the accretion processes, such as the neutron star mass, orbital eccentricity, and peculiar velocity. ECSNe should produce less massive neutron stars than CCSNe (Nomoto \cite{nomoto_1987}; Podsiadlowski et al. \cite{podsiadlowski_2004}). Thus, provided that the BeXRB subpopulations originate from different types of supernovae, we should be able to observe two subpopulations with different neutron star masses. Unfortunately, this cannot be investigated as the long orbital periods and scarcity of eclipsing BeXRBs means that there are only a handful of systems for which the neutron star mass can be determined. ECSNe are also expected to impart smaller kicks to the neutron stars and expel less matter in the explosion, producing systems with lower eccentricities and lower peculiar velocities (Podsiadlowski et al. \cite{podsiadlowski_2004}). It is expected that the observed eccentricities are still close to the original values just after the supernova explosion because the timescales for tidal circularization for the BeXRBs with orbital periods $P_{\mathrm{orb}} \geq 10$ d are significantly higher than the secondary lifetime. As a result, tidal effects should have little effect on the orbit of a typical BeXRB, which has an orbital period well above 10~d (van den Heuvel et al. \cite{heuvel_2000}; Reig \cite{reig_2011}). KCP noted that there is a trend toward lower eccentricities in the short-period BeXRB subpopulation, in line with their hypothesis, but this trend is not significant. A small number of systems for which the eccentricity values are available makes it difficult to investigate this using this parameter. 

Studying the kinematics of BeXRBs has also been problematic. Many of them have only a weak optical counterpart and/or lie at a considerable distance. Thus, their distances and proper motions have been unreliable, often with different astrometric catalogs giving disparate values of the parallaxes and proper motions for the same system (e.g., Ankay et al. \cite{ankay_2001}; Gvaramadze et al. \cite{gvaramadze_2011}). This has made the determination of their peculiar velocities difficult and viable only for the close systems. The situation has changed with the advent of the second Gaia data release (GDR2;  Gaia collaboration \cite{gaia_gdr2_summary}; Lindegren et al. \cite{lindegren_2018}), which contains the parallaxes and proper motions for the majority of the confirmed and candidate BeXRBs in the Galaxy. This enables us to derive the peculiar velocities of the Galactic population of BeXRBs in a consistent way. However, at the time of writing, the kinematics of BeXRBs situated in the Magellanic Clouds still has to be studied indirectly.

In this work, we investigate whether kinematic subpopulations of BeXRBs exist and what their origin is. We compute the tangential peculiar velocities of Galactic  BeXRBs and conduct statistical tests to determine whether they comprise two distinct subpopulations. For the population in the Small Magellanic Cloud (SMC), we use the distances from the closest star cluster that is assumed to be the birthplace of that particular BeXRB as a proxy for the tangential peculiar velocity and conduct equivalent tests. 
\section{Milky Way Be/X-ray binary population}

To construct our sample of Galactic BeXRBs, we selected the systems classified as such in Walter et al. (\cite{walter_2015}) and the systems listed in the fourth\textsuperscript{} edition of the Catalogue of High Mass X-ray binaries in the Galaxy (Liu et al. \cite{liu_2008}). To select a population of the confirmed BeXRBs, we only selected sources with measured $P_{\mathrm{s}}$. Sources without an optical counterpart or those without a match in GDR2 were also removed from the analysis. This selection yielded 24 systems out of the total of 32 listed in Walter et al. (\cite{walter_2015}). Another 11 sources satisfying the same criteria were added from Liu et al. (\cite{liu_2008}). We chose to discard \object{4U 1901+03} because its optical counterpart is unknown (Reig \& Milonaki \cite{reig_2016}; Walter et al. \cite{walter_2015}), although it is  matched to \object{Gaia DR2 4268774695647764352} in Simbad\footnote{\url{http://simbad.u-strasbg.fr/simbad/}}. We also removed \object{SAX J0635.2+0533} because of its rotation-powered (rather than accretion-powered) nature (La Palombara \cite{la_palombara_2017}). The relevant properties of the 33 selected sources are summarized in Table~\ref{table_1} without these two systems. Because we relied on the $P_{\mathrm{s}}$ values to divide the subpopulations, we refer to them as the short-spin subpopulation and the long-spin subpopulation.

We emphasize here that we only studied the tangential (transverse) components of the peculiar velocities. To obtain complete information about the kinematics of the source, the radial component of the velocity is necessary as well. However, there are literature radial velocity measurements for only a handful of the sources from the sample, and \textit{Gaia} does not provide radial velocity measurements for the early-type stars. Moreover, the measured radial velocities of OB stars are, in general, not accurate because the optical lines are formed in atmospheric layers that have outflow velocities of 20-30 $\mathrm{km/s}$ (e.g., van Oijen \cite{van_oijen_89}). We show below that this value is comparable to the typical tangential peculiar velocities of the BeXRB population. It is also not feasible to correct for this effect because the outflow velocities are variable: nonradial pulsations and wind fluctuations change them. This means that the measured radial velocities of BeXRB do not reflect the true radial motion of the system. We therefore did not consider the radial velocities and assumed that the peculiar velocity distribution of BeXRBs is isotropic. 

Many possible methods can be employed to study the kinematics of BeXRBs. First, it is possible to measure the peculiar velocities by accounting for and removing the kinematic components stemming from the Galactic rotation and the peculiar movement of the Sun relative to its local standard of rest (i.e., relative to the expected motion in the Galaxy; e.g., van den Heuvel et al. \cite{heuvel_2000}, Gvaramadze et al. \cite{gvaramadze_2011}). This method was adopted to compute the peculiar tangential velocities in this paper. More precise and accurate results might be obtained if it were possible to establish the birthplace of the studied system, such as the parent star cluster or association. In this case, the birthplace can be used to anchor a local standard of rest of the system, and the peculiar velocity of the system can then be obtained by subtracting the proper motion of the birthplace from the proper motion of the system (Ankay et al. \cite{ankay_2001}; Drew et al. \cite{drew_2018}; Lennon et al. \cite{lennon_2018}; Kalari et al. \cite{kalari_2019}). This normally results in more precise and accurate results than the previous method, as it is not dependent on us knowing the Galactic rotation curve, the distance to the Galactic center, and the peculiar movement of the Sun, and thus is not affected by the uncertainties in these parameters. However, the census of Galactic open clusters and associations beyond 1 -- 2 kpc is incomplete, particularly because of high interstellar extinction in the Galactic plane where the majority of these objects are located. This is also evident considering the high number of newly discovered star clusters using the GDR2 data (e.g., Castro-Ginard et al. \cite{castro-ginard_2018}; Cantat-Gaudin et al. \cite{cantat-gaudin_2018}; Liu \& Pang \cite{liu_2019}). Because most of the sample BeXRBs lie at much larger distances, using this method was not practical for the vast majority of sources and therefore was not used. Another possible method to obtain an estimate of the peculiar velocity is to estimate the local standard of rest by averaging the proper motions of the stars that are close in projection on the sky to the studied source and approximately at the same distance as the studied source (Kochanek et al. \cite{kochanek_2019}). Similarly, the resulting peculiar velocity estimate can then be obtained by subtracting this mean local proper motion from the proper motion of the studied source. 

An alternative way for studying the kinematical properties of the BeXRB sample is indirectly, by studying the locations of BeXRBs and the sites of recent massive star formation, such as the young open clusters and associations. On the premise that the closest young cluster or association to the particular BeXRB system is its birthplace, the separation between the two can then serve as a proxy for the peculiar velocity. This method was used to study the kinematics of BeXRBs in the SMC (Coe \cite{coe_2005}), but it is ill-suited for the Milky Way systems because the cluster and association catalogs are incomplete, as mentioned above. 

\subsection{Peculiar velocities using the Bailer-Jones scale length}

Going from the noisy parallax and proper motion measurements to distances and velocities is non-trivial. The most notable problems are the nonlinearity of the transformation and the positivity constraint of the distance. The parallax measurements often exhibit high relative uncertainties and can even be negative, which is often the case when distant objects such as BeXRBs are considered. The naive methods fail and give unphysical results when these measurements are used. However, these measurements are perfectly valid and still hold informational value, therefore it would be a mistake to discard them. The only viable way to handle these measurements is to use a probabilistic analysis (see Luri et al. \cite{luri_2018} and Bailer-Jones et al. \cite{bailer-jones_2018} for a more detailed discussion).  

Here we used the parallaxes and proper motions from GDR2 (Gaia collaboration \cite{gaia_gdr2_summary}; Lindegren et al. \cite{lindegren_2018}) to compute the tangential peculiar velocities. We followed the approach outlined in Luri et al. (\cite{luri_2018}), using the workflow from Bailer-Jones (\cite{bailer-jones_workflow}). The distances and tangential velocities were jointly estimated from the parallaxes and proper motions by Bayesian inference, with the prior scale lengths for each object adopted from Bailer-Jones et al. (\cite{bailer-jones_2018}). To obtain the peculiar tangential velocities, we adopted the solar Galactocentric distance $R_{0} = 8.2$ kpc, the circular Galactic rotation velocity $\Theta_{0} = 238\, \mathrm{km}\, \mathrm{s}^{-1}$ , and the solar peculiar motion $(U_{\odot}, V_{\odot}, W_{\odot}) = (10.0, 11.0, 7.0)\, \mathrm{km}\, \mathrm{s}^{-1}$ from Bland-Hawthorn \& Gerhard (\cite{bland-hawthorn_2016}). 

\subsection{Peculiar velocities using an empirically determined scale length}

BeXRBs are predominantly discovered in X-rays, which means that it is possible to detect them at significant distances and/or obscured by several tens of magnitudes of extinction. They are then subject to deep follow-up observations to determine and characterize the stellar counterpart. This may result in their selection function being different from the other field stars. However, we do not expect this to affect our results as significantly as in Gandhi et al. (\cite{gandhi_2018}), who studied low-mass X-ray binaries hosting black holes. BeXRBs have significantly shorter lifetimes, therefore it is not possible for them to escape far from their parent populations near the Galactic plane, even at runaway speeds. Considering a runaway velocity of 30 $\mathrm{km}\, \mathrm{s}^{-1}$ $\sim$ 30 $\mathrm{pc}\, \mathrm{Myr}^{-1}$ (informed by the measured tangential peculiar velocities of BeXRBs by van den Heuvel et al. \cite{heuvel_2000} and theoretical peculiar velocity predictions from Eldridge et al. \cite{eldridge_2011} and Renzo et al. \cite{renzo_2019}) and a BeXRB lifetime after the supernova of 10~Myr, we obtain a migration distance estimate of 300 pc. There is also evidence that BeXRB progenitors tend to remain bound within their parent cluster or associations and only acquire high peculiar velocities later on after the supernova explosion of one of its components (Bodaghee et al. \cite{bodaghee_2012}). This further limits the possible displacement from the Galactic-plane massive star population. On the other hand, Treuz et al. (\cite{treuz_2018}) found that there might be problems with the distances determined using the scale lengths from Bailer-Jones et al. (\cite{bailer-jones_2018}), where these distances appear to be an underestimation when compared to the distance values obtained using the conventional methods for sources closer than $\sim$ 5 kpc. For sources that lie farther away than this, the trend seems to be reversed.

\begin{table*}
\caption{Galactic BeXRB pulsars. Names, orbital periods ($P_{\mathrm{orb}}$), and spin periods ($P_{\mathrm{s}}$) are obtained from Liu et al. (\cite{liu_2008}) and Walter et al. (\cite{walter_2015}). GDR2 ID is the source id of the counterpart in GDR2.  $d_{\mathrm{lit}}$ are the distances of the BeXRBs collected from the literature used to estimate the scale length (Sect. 2.2). The typical uncertainty for $P_{\mathrm{orb}}$ and $P_{\mathrm{s}}$ is $<$ 1~d and $\ll$ 1~s, respectively (see references in Liu et al. \cite{liu_2008} and Walter et al. \cite{walter_2015}), although higher errors may be present for higher $P_{\mathrm{orb}}$ and $P_{\mathrm{s}}$ values.}      
\label{table_1}      
\centering          
\begin{tabular}{lllllll}
\hline\hline       
Name    & $P_{\mathrm{orb}}$  & $P_{\mathrm{s}}$  & GDR2 ID & $d_{\mathrm{lit}}$   & $d_{\mathrm{BJ}}$ &  $d_{\mathrm{lit}}$ ref      \\
 & (d) & (s) & & (kpc) &  (kpc) & \\
\hline                    
 \object{4U 0115+63}            & 24.3    & 3.61      & 524677469790488960  & $5.3 \pm 0.44$  & $7.2_{-1.1}^{+1.5}$  &  Coleiro \& Chaty (\cite{coleiro_chaty_2013}) \\
 \object{V 0332+53}             & 34.67   & 4.375      & 444752973131169664  & $6.9 \pm 0.71$ & $5.13_{-0.76}^{+1}$ &  Coleiro \& Chaty (\cite{coleiro_chaty_2013}) \\
 \object{GS 0834-430}           & 105.8   & 12.3       & 5523448270462666880\tablefootmark{a} & $3.0<d<5.0$ & $5.5_{-1.7}^{+2.5}$  & Israel et al. ( \cite{israel_2000}) \\
 \object{IGR J19294+1816}       & 117.2   & 12.4       & 4323316622779495680 & $11.0 \pm 1.0 $ & $2.93_{-1.5}^{+2.5}$   &  Rodes-Roca et al. (\cite{rodes-roca_2018}) \\
 \object{XTE J1946+274}         & 169.2   & 15.8       & 2028089540103670144 & $6.2 \pm 3.0 $ & $12.6_{-2.9}^{+3.9}$  &  Coleiro \& Chaty (\cite{coleiro_chaty_2013}) \\
 \object{4U 1416-62}            & 42.12   & 17.64     & 5854175187680510336\tablefootmark{b} & $7.0 \pm 0.74 $  & $5.21_{-1.6}^{+2.6}$  &  Coleiro \& Chaty (\cite{coleiro_chaty_2013}) \\
 \object{KS 1947+300}           & 40.415  & 18.7      & 2031939548802102656 & $8.5 \pm 2.3 $ & $15.2_{-2.7}^{+3.7}$  &  Coleiro \& Chaty (\cite{coleiro_chaty_2013}) \\
 \object{GS 1843+00}            & & 29.5      & 4278536022438800640 &  $12.5 \pm 2.5 $ & $2.28_{-0.95}^{+1.9}$  &  Israel et al. (\cite{israel_2001}) \\
 \object{RX J0812.4-3114}       & 81.3    & 31.8851    & 5548261400354128768 & $8.6 \pm 1.8 $  & $6.76_{-0.91}^{+1.2}$  &  Coleiro \& Chaty (\cite{coleiro_chaty_2013}) \\
 \object{EXO 2030+375}          & 46.016  & 42        & 2063791369815322752 & $3.1 \pm 0.38 $ & $3.64_{-0.88}^{+1.3}$ & Coleiro \& Chaty (\cite{coleiro_chaty_2013})  \\
 \object{IGR J22534+6243}       &   & 46.67     & 2207277877757956352 & $4.0<d<5.0$ & $8.06_{-1.6}^{+2.3}$  & Esposito et al. (\cite{esposito_2013}) \\
 \object{AX J1700.2-4220}       & 44.0    & 54        & 5966213219190201856 & $1.7<d<2.6$ & $1.56_{-0.14}^{+0.18}$   & Negueruela \& Schurch (\cite{negueruela_2007}) \\
 \object{Cep X-4}               &      & 66.2       & 2178178409188167296 & $3.7 \pm 0.52 $ & $10.2_{-1.6}^{+2.1}$  &  Coleiro \& Chaty (\cite{coleiro_chaty_2013}) \\
 \object{XTE J1906+090}         & 26--30  & 89.17       & 4310649149314811776\tablefootmark{c} & $d > 4$  & $2.77_{-1.4}^{+2.3}$  & G{\"o}{\v{g}}{\"u}{\c{s}} et al.  (\cite{gogus_2005}) \\
 \object{GRO J1008-57}          & 249.46  & 93.6       & 5258414192353423360\tablefootmark{d} & $4.1 \pm 0.59 $  & $3.65_{-0.4}^{+0.51}$  & Coleiro \& Chaty (\cite{coleiro_chaty_2013}) \\
 \object{3A 0535+262}           & 111.1   & 103       & 3441207615229815040 & $3.8 \pm 0.33 $ & $2.13_{-0.21}^{+0.26}$ & Coleiro \& Chaty (\cite{coleiro_chaty_2013}) \\
 \object{4U 0728-25}            & 34.5    & 103.2      & 5613494119544761088 & $5.0 \pm 0.82 $ & $9.51_{-2.1}^{+3.1}$  &  Coleiro \& Chaty (\cite{coleiro_chaty_2013}) \\
 \object{2E 0655.8-0708}        & 101.2   & 160.7     & 3052677318793446016 & $3.9 \pm 0.1 $ & $5.11_{-0.93}^{+1.4}$ &  McBride et al.( \cite{mcbride_2006}) \\
 \object{IGR J11435-6109}       & 52.46   & 161.76     & 5335021664274920576 & $9.8 \pm 0.86 $ & $8.59_{-1.8}^{+2.5}$   &  Coleiro \& Chaty (\cite{coleiro_chaty_2013}) \\
 \object{GRO J2058+42}          & 55.03   & 198        & 2065653598916388352 & $9.0 \pm 1.3 $ & $8.04_{-0.94}^{+1.2}$  & Wilson et al. (\cite{wilson+2005}) \\
 \object{RX J0440.9+4431}       & 155     & 202.5      & 252878401557369088  & $2.9 \pm 0.37 $ & $3.25_{-0.45}^{+0.62}$  &  Coleiro \& Chaty (\cite{coleiro_chaty_2013})  \\
 \object{GX 304-1}              & 132.5   & 272       & 5863533199843070208 & $1.3 \pm 0.1 $ & $2.01_{-0.13}^{+0.15}$ & Coleiro \& Chaty (\cite{coleiro_chaty_2013})  \\
 \object{4U 1145-619}           & 187.5   & 292       & 5334823859608495104 &  $4.3 \pm 0.52 $ & $2.23_{-0.16}^{+0.19}$   & Coleiro \& Chaty (\cite{coleiro_chaty_2013})  \\
 \object{SAX J2103.5+4545}      & 12.68   & 358.6     & 2162805896614571904 &  $8.0 \pm 0.78 $ & $6.43_{-0.69}^{+0.86}$  &  Coleiro \& Chaty (\cite{coleiro_chaty_2013}) \\
 \object{1A 1118-615}           & 24.0    & 406         & 5336957010898124160 & $3.2 \pm 1.4 $ & $2.93_{-0.22}^{+0.26}$ &  Coleiro \& Chaty (\cite{coleiro_chaty_2013}) \\
 \object{IGR J01583+6713}       & & 469.2     & 518990967445248256  &  $4.1 \pm 0.63$ & $7.4_{-0.9}^{+1.1}$   &  Coleiro \& Chaty (\cite{coleiro_chaty_2013}) \\
 \object{2RXP J130159.6-635806} & & 700        & 5862285700835092352 & $4.0<d<7.0$ & $5.54_{-1.7}^{+2.8}$ & Chernyakova et al. (\cite{chernyakova_2005})   \\
 \object{4U 0352+309}           & 250     & 835       & 168450545792009600  & $1.2 \pm 0.16 $ & $0.793_{-0.034}^{+0.037}$  & Coleiro \& Chaty (\cite{coleiro_chaty_2013}) \\
 \object{3U 1022-55}            & & 860        & 5352018121173519488 & $\sim 5$ & $5.04_{-0.75}^{+1}$ & Motch et al. (\cite{motch_97}) \\
 \object{SAX J2239.3+6116}      & 262     & 1247      & 2201091578667140352 & $4.9 \pm 0.8 $ & $8.03_{-1}^{+1.3}$  & Reig et al. (\cite{reig_2017}) \\
 \object{RX J0146.9+6121}       &  & 1400      & 511220031584305536  & $2.3 \pm 0.5$  & $2.5_{-0.2}^{+0.2}$ & Reig et al. (\cite{reig_97})  \\
 \object{4U 2206+543} &  9.57 &  5559 &  2005653524280214400 & $3.4 \pm 0.35 $ & $3.34_{-0.32}^{+0.39}$ & Coleiro \& Chaty (\cite{coleiro_chaty_2013}) \\
 \object{1H 1249-637}           &  & 14200      & 6055103928246312960 &  $0.392 \pm 0.055 $ & $0.416_{-0.021}^{+0.023}$  &  Megier et al. (\cite{megier_2009}) \\
\hline                  
\end{tabular}
\tablefoot{Counterparts:
\tablefoottext{a}{Israel et al. (\cite{israel_2000})},  \tablefoottext{b}{Grindlay et al. (\cite{grindlay_84})}, \tablefoottext{c}{G{\"o}{\v{g}}{\"u}{\c{s}} et al. (\cite{gogus_2005})}, \tablefoottext{d}{Coe et al. (\cite{coe_94})}
}
\end{table*}

To investigate this, we determined a new scale length for the selected BeXRBs using the distance measurements collected from the literature. These measurements come from a number of sources and thus were collected using various methods, sometimes applied in conjunction. For some sources, more distance estimates exist, in which case we then preferred to use the estimates from the most recent works. These estimates, together with their errors (if available), are listed in Table ~\ref{table_1}.

These distances were then fit with the exponentially decreasing space density prior probability model,
\begin{equation}
P(d) = \frac{1}{2L^{3}}d^{2}\mathrm{e}^{-d/L} \quad \quad \quad  \mathrm{if} \quad  d>0,
\end{equation}
where $d$ is the distance to the source and $L$ is the scale length, as discussed in Bailer-Jones (\cite{bailer-jones_2015}).

Similarly to Gandhi et al. (\cite{gandhi_2018}), an unbinned maximum likelihood algorithm was used for the fit. To quantify the uncertainty of $L$, we generated randomized ensembles of $d_{\mathrm{lit}}$ values by resampling from a suitable distribution for each object. In most cases, for the objects with a published distance estimate $d_{\mathrm{lit}}$ and its uncertainty, a normal distribution was used for resampling, with the assumed mean and standard deviation corresponding to $d_{\mathrm{lit}}$ and its uncertainty, respectively. For four sources with distance limits, random values were drawn from a uniform distribution characterized by the lower and upper distance limit (GS 0834-430, IGR J22534+6243, AX J1700.2-4220, and 2RXP J130159.6-635806). In the case of XTE J1906+090, where only the lower limit on $d_{\mathrm{lit}}$ is known, we also drew from a uniform distribution, where we assumed the upper limit to be $d_{\mathrm{lit}} + 5 $ kpc. One object in our sample, 3U 1022-55, has no uncertainty on $d_{\mathrm{lit}}$ . In this case, we assumed the uncertainty to be 20 \% of the published $d_{\mathrm{lit}}$. We then resampled from a normal distribution as above.

\begin{figure}
  \resizebox{\hsize}{!}{\includegraphics{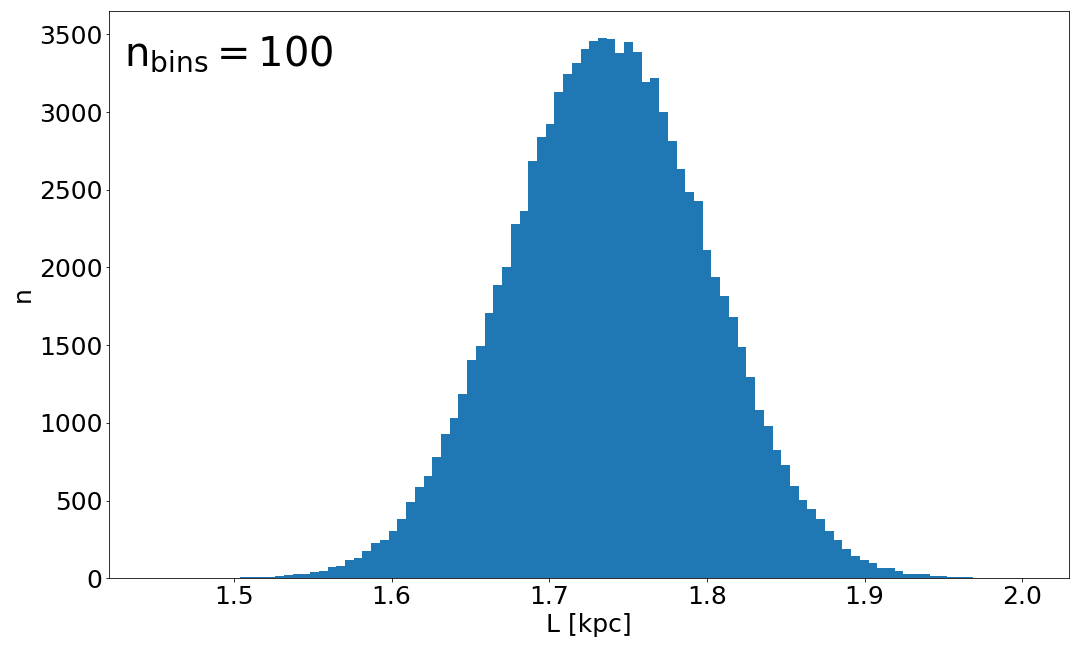}}
  \caption{Histogram of the scale lengths $L$ resulting from the resampling.}
  \label{sl_histogram}
\end{figure}

A total of 100 000 ensembles were randomized, resulting in a mean value of the characteristic scale length of $L = 1.74~\pm~0.06$~kpc (this is somewhat higher than the scale height of the thick disk of 0.7--1.2 kpc; Siegel et al. \cite{siegel_2002}), with the uncertainty quoted here being the standard deviation of the randomized ensambles. The scale length distribution is plotted in Fig~\ref{sl_histogram}. We then adopted this scale length for all the objects in our sample and followed the same procedure as outlined in Sect. 2.1.

\subsection{Quality of GDR2 data and astrometric fits}
While GDR2 provides astrometric measurements of unprecedented quality and quantity, which allows insight into the kinematics of many BeXRBs for the first time, it still contains some sources for which the solutions are ill-behaved. The astrometric parameters of these sources should be considered suspect and be flagged or filtered out from the analysis. Problems with the astrometry may arise for sources that are located in regions with high source densities, for instance, in the Galactic plane. Another potential caveat is that all GDR2 sources are treated as single stars in the astrometric solution, where binaries do not receive any special treatment (Gaia Collaboration et al. \cite{gaia_gdr2_summary}, Lindegren et al. \cite{lindegren_2018}). Owing to the binary and early-type nature of BeXRBs, they might be affected by this problem. It is therefore necessary to examine the quality of the astrometric parameters of the BeXRB sample. 

Unreliable astrometric solutions can be empirically identified by considering the distributions of the parallax and proper motion errors at the relevant magnitudes and colors and comparing them to the errors for the objects of interest, or by using the recommended astrometric quality indicators that are included or can be computed from the parameters that are part of GDR2. Informed by Gaia Collaboration et al. (\cite{gaia_gdr2_summary}), Lindegren et al. (\cite{lindegren_2018}), and Lindegren (\cite{lindegren_2018_tn}), we retained the systems that satisfied 

\noindent
\begin{tt}
\begin{itemize}
\item duplicated\_source = False
\item astrometric\_excess\_noise < 1 mas or astrometric\_excess\_noise\_sig < 2
\item ruwe < 1.4 or u < $1.2\times \max(1, \exp (-0.2(G -19.5))).$
\end{itemize}
\end{tt}
The flag {\tt duplicated\_source=True} indicates observational, cross-matching, or processing problems, or stellar multiplicity, probably leading to problems in the astrometric solution. The {\tt astrometric\_excess\_noise} ($\epsilon_{i}$) is an angular measure of the astrometric goodness of fit, indicating the additional scatter that may arise from the movement of the emission centroid that in turn is due to the motion of the components inside a binary. Last, the cuts based on {\tt ruwe} and {\tt u}, which stand for renormalized unit weight error and unit weight error, respectively, ensured the removal of ill-behaved astrometric solutions. The {\tt ruwe} is obtained by dividing {\tt u} by a normalization factor that is a function of the source magnitude and color and is included in the GDR2 archive.  The {\tt u} values are obtained through
\begin{equation}
u = \sqrt{\chi^{2}/(N-5)},
\end{equation}
where $\chi^{2}$ is {\tt astrometric\_chi2\_al} and $N$ is {\tt astrometric\_n\_good\_obs\_al}, which can both be queried in the GDR2 archive. Cuts based on {\tt u} were applied to the objects with no color information in GDR2, for which the {\tt ruwe} values were not available.

\begin{table*}
\caption{Relevant GDR2 parameters and flags pertaining to the quality of the astrometric solution.}  
\label{gdr2_flags}      
\centering  
\begin{tabular}{lllllllll}
\hline\hline   
system & parallax & parallax\_error & Gmag & $\epsilon_{i}$ & $\epsilon_{i}$ sig & u & ruwe & duplicated\_source \\
 & (mas) & (mas) & (mag) & (mas) & & & & \\
\hline
4U 0115+63  & 0.091 & 0.027 & 14.44 & 0.13 & 3.51 & 1.20 & 1.00 & False \\
V 0332+53  & 0.14 & 0.04 & 14.22 & 0.20 & 8.28 & 1.40 & 1.01 & False \\
GS 0834-430  & -0.16 & 0.15 & 20.52 & 0.75 & 5.96 & 1.34 & 0.96 & False \\
IGR J19294+1816  & -0.38 & 1.06 & 20.39 & 4.47 & 5.55 & 1.36 & 1.23 & False \\
XTE J1946+274  & -0.072 & 0.044 & 15.71 & 0.25 & 6.88 & 1.27 & 0.96 & False \\
4U 1416-62  & 0.0046 & 0.1352 & 17.77 & 0.33 & 2.13 & 1.10 & -- & False \\
KS 1947+300  & 0.0056 & 0.0189 & 13.84 & 0 & 0 & 0.78 & 0.96 & False \\
GS 1843+00  & 0.41 & 0.35 & 18.68 & 1.06 & 4.0 & 1.26 & -- & False \\
RX J0812.4-3114  & 0.10 & 0.02 & 12.48 & 0 & 0 & 1.27 & 1.02 & False \\
EXO 2030+375  & 0.15 & 0.11 & 16.91 & 0.64 & 13.43 & 1.70 & 1.05 & False \\
IGR J22534+6243  & 0.053 & 0.037 & 14.60 & 0.23 & 8.53 & 1.35 & 1.03 & False \\
AX J1700.2-4220  & 0.62 & 0.06 & 8.68 & 0 & 0 & 1.08 & 0.87& False  \\
Cep X-4  & 0.051 & 0.020 & 13.82 & 0 & 0 & 0.92 & 1.08 & False \\
XTE J1906+090  & 0.066 & 0.726 & 19.73 & 2.85 & 9.35 & 1.38 & -- & False \\
GRO J1008-57  & 0.24 & 0.03 & 13.90 & 0.25 & 13.41 & 1.54 & 1.02 & False \\
3A 0535+262  & 0.44 & 0.05 & 8.68 & 0 & 0 & 1.34 & 1.05 & False \\
4U 0728-25  & 0.028 & 0.039 & 11.62 & 0 & 0 & 1.23 & 0.94 & True \\
2E 0655.8-0708  & 0.15 & 0.04 & 12.03 & 0 & 0 & 1.19 & 1.10 & False \\ 
IGR J11435-6109  & 0.03 & 0.04 & 15.67 & 0.10 & 0.90 & 1.07 & 0.99 & False \\
GRO J2058+42  & 0.077 & 0.018 & 14.19 & 0 & 0 & 0.99 & 1.06 & False \\
RX J0440.9+4431  & 0.27 & 0.05 & 10.43 & 0 & 0 & 1.00 & 0.80 & False \\
GX 304-1  & 0.47 & 0.03 & 12.65 & 0 & 0 & 1.69 & 0.99 & True \\
4U 1145-619  & 0.42 & 0.04 & 8.63 & 0 & 0 & 1.33 & 0.88 & False \\
SAX J2103.5+4545  & 0.12 & 0.02 & 13.0 & 0 & 0 & 0.90 & 1.08 & False \\
1A 1118-615  & 0.31 & 0.03 & 11.60 & 0 & 0 & 1.19 & 0.94 & False \\
IGR J01583+6713  & 0.098 & 0.018 & 13.70 & 0 & 0 & 0.95 & 0.97 & False \\
2RXP J130159.6-635806 & 0.063 & 0.108 & 17.34& 0.52 & 6.66 & 1.28 & 0.95 & False \\
4U 0352+309  & 1.23 & 0.06 & 6.25 & 0.16 & 16.39 & 2.29 & 1.32 & False \\
3U 1022-55  & 0.16 & 0.03 & 11.25 & 0 & 0 & 1.55 & 1.06 & False \\
SAX J2239.3+6116  & 0.084 & 0.019 & 14.15 & 0.08 & 1.17 & 1.18 & 1.11 & False \\
RX J0146.9+6121  & 0.37 & 0.03 & 11.21 & 0 & 0 & 1.70 & 1.21 & False \\
4U 2206+543  & 0.27 & 0.03 & 9.74 & 0 & 0 & 1.44 & 0.90 & False \\
1H 1249-637  & 2.38 & 0.13 & 5.12 & 0.62 & 148.90 & 4.82 & 0.99 & False \\
\hline
\end{tabular}
\end{table*}

The relevant quantities and quality flags for the objects in the Galactic BeXRB sample are listed in Table~\ref{gdr2_flags}. Based on these quality cuts, we discarded five objects from the subsequent analysis. Table~\ref{gdr2_flags} shows that the unmodeled orbital motion due to the binary nature of the BeXRBs does not seem to affect the astrometric solutions for the majority of sources in a significant way. The longest $P_{\mathrm{orb}}$ in the sources is about 262 d for SAX J2239.3+6116 (in't Zand et al. \cite{int_zand_2000}), which has an astrometric solution well below all considered cut limits. This $P_{\mathrm{orb}}$ is still considerably shorter than the 22-month observing time of GDR2. For periods such as this and shorter, any orbital motion will therefore largely average out (Jennings et al.  \cite{jennings_2018}).

The five discarded sources (4U 0728-25, GX 304-1, GS 1843+00,  XTE J1906+090, and IGR J19294+1816) are not outstanding in the BeXRB sample in $P_{\mathrm{orb}}$, distance, or optical brightness. The last three sources, which were discarded due to the increased $\epsilon_{i}$, lie relatively close to each other in the same region of the sky. They also have lower values of {\tt visibility\_periods\_used}, ranging from 10 to 13, while the mean value for the studied BeXRB sample is 15. We opted to list the computed velocities of these sources in the subsequent tables, but we did not consider them in the statistical analysis as their velocities cannot be considered reliable.

\subsection{Kinematics of Galactic BeXRBs}

\begin{figure}
  \resizebox{\hsize}{!}{\includegraphics{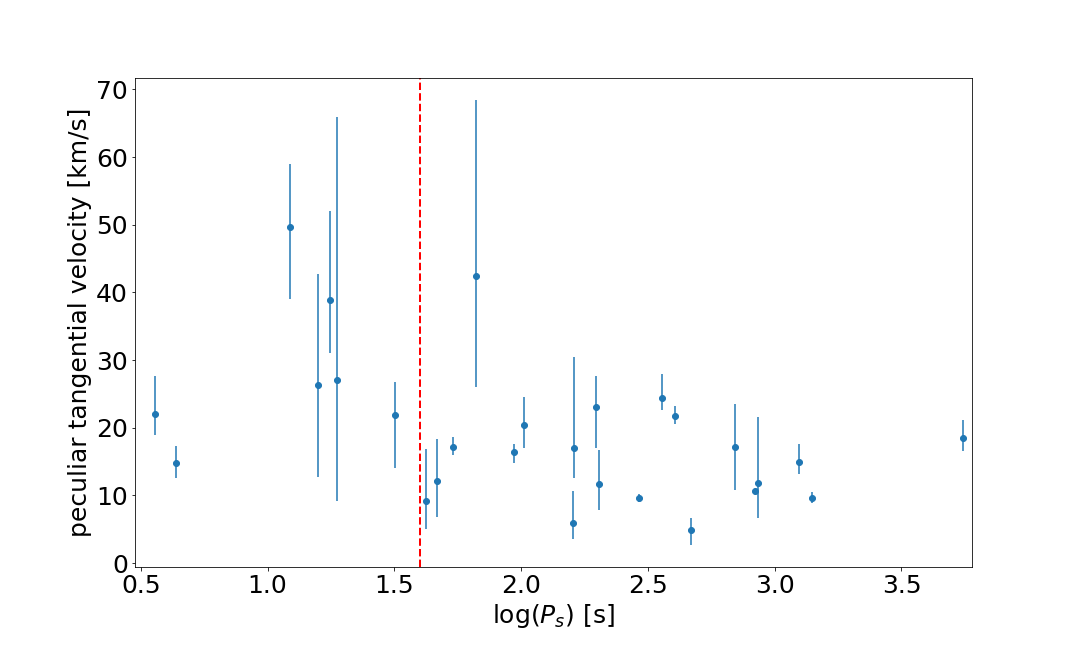}}
 \resizebox{\hsize}{!}{\includegraphics{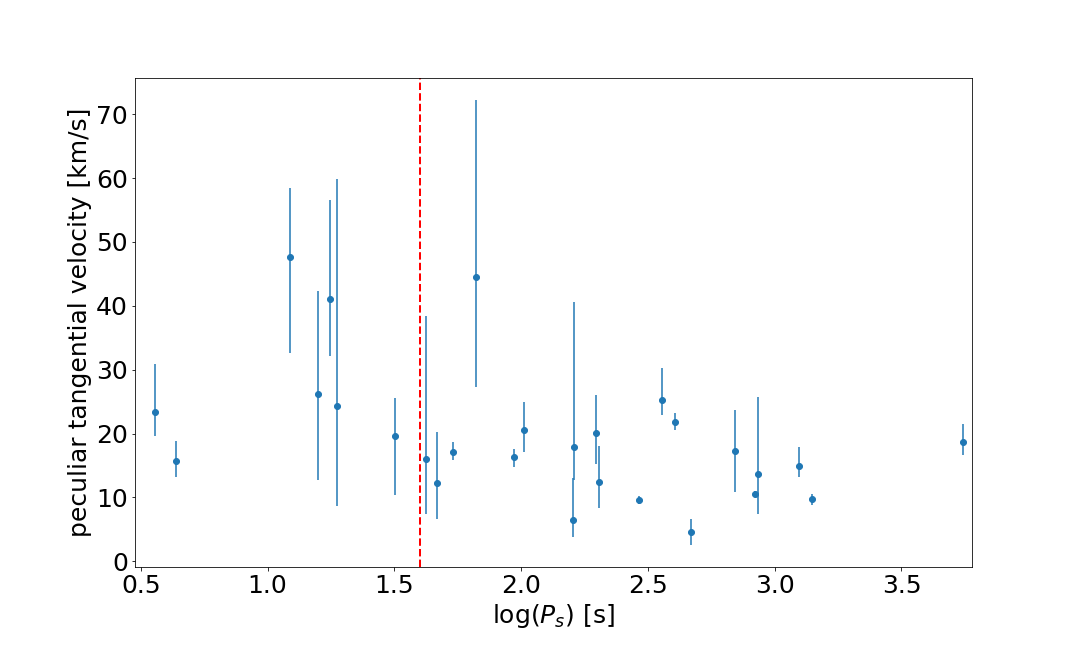}}
  \caption{Peculiar tangential velocities of Galactic BeXRBs with respect to $P_{\mathrm{s}}$. The red vertical line delineates the approximate divide ($P_{\mathrm{s, split}}=$ 40 s) between the two BeXRB subpopulations according to KCP. Top: Tangential peculiar velocities computed using the scale lengths from Bailer-Jones et al. (\cite{bailer-jones_2018}). Bottom: Same as above, but with the the scale length $L$ resulting from the resampling we used.}
  \label{MWX_gaiadr2}
\end{figure}

 We obtained the peculiar velocities for 33 Galactic BeXRBs showing pulsations (see Table~\ref{table_2}). In addition to the five sources we discarded because the astrometry was unreliable, we also decided to remove 1H 1249-637 because of its uncertain nature as a $\mathrm{\gamma}$ Cas analog (these are systems that most likely do not host a neutron star, where the X-ray emission is generated by interactions between magnetic fields on the Be star and its decretion disk; e.g., Smith et al. \cite{smith_2016}), yielding a final sample of 27 sources. However, this did not affect the results of the following analysis in any significant way. We split the sample according to the dip in  $P_{\mathrm{s}}$ adopted from KCP, which is $P_{\mathrm{s, split}}=$~40~s. The short-spin subpopulation comprises 7 sources, and the long-spin subpopulation is more numerous, with 20 sources. The distribution of the tangential peculiar velocities, derived using the scale lengths from Bailer-Jones et al. (\cite{bailer-jones_2018}; $v_{\mathrm{pec, BJ}}$) and the empirically determined scale length from Sect. 2.2 ($v_{\mathrm{pec, iso}}$), with respect to $P_{\mathrm{s}}$ is shown in Fig.~\ref{MWX_gaiadr2}.

\begin{table*}
\caption{Derived tangential peculiar velocities and orbital eccentricity values compiled from the literature. $v_{\mathrm{pec, BJ}}$ denotes the velocities obtained using the priors adopted from Bailer-Jones et al. (\cite{bailer-jones_2018}), and $v_{\mathrm{pec, iso}}$ are the velocities obtained using the empirically determined scale length derived in Sec. 2.2.}      
\label{table_2}      
\centering  
\begin{tabular}{lllll}
\hline\hline       
Name    & $v_{\mathrm{pec, BJ}}$   & $v_{\mathrm{pec, iso}}$ & $e$  & $e$ ref \\
 & ($\mathrm{km}\, \mathrm{s}^{-1}$) & ($\mathrm{km}\, \mathrm{s}^{-1}$) \\
\hline        
4U 0115+63            & $22_{-3}^{+6}$ & $23_{-4}^{+8}$  & $0.342 \pm 0.004 $& Raichur \& Paul (\cite{raichur_paul_2010}) \\
V 0332+53             & $15_{-2}^{+3}$  & $16_{-2}^{+3}$ & $0.417 \pm 0.007$ & Raichur \& Paul (\cite{raichur_paul_2010}) \\
GS 0834-430           & $50_{-11}^{+9}$  & $48_{-15}^{+11}$ & $0.14 \pm 0.04$ & Wilson et al. (\cite{wilson_97}) \\
IGR J19294+1816       & $73_{-45}^{+81}$ & $99_{-60}^{+93}$ & &  \\
XTE J1946+274         & $26_{-14}^{+17}$ &$26_{-14}^{+16}$  & $0.33 \pm 0.05$ & Wilson et al. (\cite{wilson_2003})  \\
4U 1416-62            & $39_{-8}^{+13}$ & $41_{-9}^{+16}$ & $0.417 \pm 0.003$ & Raichur \& Paul (\cite{raichur_paul_2010})  \\
KS 1947+300           & $27_{-18}^{+39}$ & $24_{-16}^{+36}$ & $0.033 \pm 0.013$ & Galloway et al. (\cite{galloway_2004}) \\ 
GS 1843+00            & $17_{-9}^{+16}$ & $23_{-13}^{+65}$ & &  \\
RX J0812.4-3114       & $22_{-8}^{+5}$ & $20_{-9}^{+6}$ & &  \\
EXO 2030+375          & $9_{-4}^{+8}$  & $16_{-9}^{+22}$ & $0.412 \pm 0.001$ & Wilson et al. (\cite{wilson_2008}) \\
IGR J22534+6243       & $12_{-5}^{+6}$  & $12_{-6}^{+8}$  & & \\
AX J1700.2-4220       & $17_{-1}^{+1}$  & $17_{-1}^{+2}$  & & \\
Cep X-4               & $42_{-16}^{+26}$ & $45_{-17}^{+28}$  & & \\
XTE J1906+090         & $51_{-36}^{+79}$  & $86_{-60}^{+98}$ & $0.03 < e < 0.06$ & Wilson et al. (\cite{wilson_2002}) \\
GRO J1008-57          & $16_{-2}^{+1}$ & $16_{-2}^{+1}$  & $0.68 \pm 0.02$ & Coe et al. (\cite{coe_2007}) \\
3A 0535+262           & $20_{-3}^{+4}$  & $21_{-3}^{+4}$ & $0.47 \pm 0.02$ & Finger et al. (\cite{finger_1994}) \\
4U 0728-25            & $13_{-8}^{+10}$  & $13_{-8}^{+10}$ & & \\
2E 0655.8-0708        & $5.9_{-2.3}^{+4.7}$  & $6.5_{-2.7}^{+6.6}$  & $e \sim 0.4$ & Yan et al. (\cite{yan_2012}) \\
IGR J11435-6109       & $17_{-5}^{+14}$ & $18_{-5}^{+23}$ & & \\
GRO J2058+42          & $23_{-6}^{+5}$   & $20_{-5}^{+6}$ & & \\
RX J0440.9+4431       &  $12_{-4}^{+5}$ & $12_{-4}^{+6}$ & $e > 0.4$ & Yan et al. (\cite{yan_2016}) \\
GX 304-1              &  $23_{-1}^{+1}$ & $23_{-1}^{+1}$  &  $e \sim 0.5$ & Sugizaki et al. (\cite{sugizaki_2015}) \\
4U 1145-619           & $9.6_{-0.6}^{+0.6}$  & $9.6_{-0.6}^{+0.6}$  & $e \sim 0.8$ & Watson et al. (\cite{watson_1981}) \\
SAX J2103.5+4545      & $24_{-2}^{+4}$ & $25_{-2}^{+5}$  & $0.406 \pm 0.004$ & Baykal et al. (\cite{baykal_2007}) \\
1A 1118-615           & $22_{-1}^{+1}$  & $22_{-1}^{+1}$  & $0 < e < 0.16$ & Staubert et al. (\cite{staubert_2011}) \\
IGR J01583+6713       & $4.8_{-2.1}^{+1.9}$  & $4.7_{-2.1}^{+2.0}$  & & \\  
2RXP J130159.6-635806 & $17_{-6}^{+6}$  & $17_{-6}^{+6}$ & & \\
4U 0352+309           & $10.6_{-0.4}^{+0.4}$ & $10.6_{-0.4}^{+0.4}$ & $0.111 \pm 0.018$ & Delgado-Mart{\'\i} et al. (\cite{delgado-marti_2001}) \\
3U 1022-55            & $12_{-5}^{+10}$ & $14_{-6}^{+12}$ & & \\
SAX J2239.3+6116      & $15_{-2}^{+3}$ & $15_{-2}^{+3}$  & & \\
RX J0146.9+6121       & $9.6_{-0.8}^{+0.9}$  & $9.7_{-0.8}^{+0.9}$  & & \\
4U 2206+543           & $18_{-2}^{+3}$  & $19_{-2}^{+3}$ & $e \sim 0.15$ & Rib{\'o} et al. (\cite{ribo_2006}) \\
1H 1249-637           & $2.2_{-0.8}^{+1.0}$  & $2.1_{-0.8}^{+1.0}$  & & \\
\hline
\end{tabular}
\end{table*}

Although only a few BeXRBs fall into the short-spin subpopulation, it is apparent that this subpopulation seems on average to be moving with a higher peculiar velocity than the long-spin subpopulation. Using the means of the tangential peculiar velocity posteriors of the individual BeXRBs, we can estimate the characteristics of the two BeXRB subpopulations. The short-spin subpopulation possesses a mean tangential peculiar velocity of approximately 29$\pm$11 $ \mathrm{km}\, \mathrm{s}^{-1}$, which  for the long-spin subpopulation is about 16$\pm$8 $ \mathrm{km}\, \mathrm{s}^{-1}$. The reported errors are the standard deviations of the velocity mean distributions. The results obtained using the empirically determined scale length derived in Sect 2.2 are similar, but individual velocity measurements exhibit higher uncertainties. In this case, we obtained 28$\pm$11 $ \mathrm{km}\, \mathrm{s}^{-1}$ and 16$\pm$8 $ \mathrm{km}\, \mathrm{s}^{-1}$ for the short- and long-spin subpopulation, respectively.

 To test the significance of this difference between the populations, we conducted a two-sample Anderson-Darling test (e.g., Scholz \& Stephens \cite{anderson_darling}). The result is that the two populations are indeed distinct. We also used bootstrap testing to estimate the velocity difference between the populations and quantified the effect size through Cohen's d (Cohen \cite{cohen_1988}), which is the difference in subpopulation means $\bar{X}_{1}$ and $\bar{X}_{2}$, standardized by dividing by the standard deviation:
\begin{equation}
d_{C} = \frac{\bar{X}_{1} - \bar{X}_{2}}{S},
\end{equation}
where $S$ is the pooled standard deviation:
\begin{equation}
S = \sqrt{\frac{(n_{1} - 1)s_{1}^{2} + (n_{2} - 1)s_{2}^{2}}{n_{1} + n_{2} - 2}},
\end{equation}
where $n_{i}$ and $s_{i}$ are the size and standard deviation of the subpopulation $i$, respectively. The results of this testing are summarized in Table~\ref{statistics_MW_BeXRBs}.

\begin{table}
\caption{Summary of the statistical properties of the Galactic BeXRB sample. $p$ indicates the significance of the subpopulation split as obtained from the Anderson-Darling test,  pop. $v_{\mathrm{pec}}$ difference is the difference between the short-spin and and long-spin subpopulation velocity means obtained from bootstrapping, and $d_{C}$ is the value of Cohen's d, indicating the effect size. The difference between the subpopulation velocity means, pop. $v_{\mathrm{pec}}$, is characterized using a 95\% credibility interval as resulting from the bootstrap testing.}           
\label{statistics_MW_BeXRBs}      
\centering                          
\begin{tabular}{l l l }        
\hline\hline                 
 & short $P_{\mathrm{s}}$ pop. & long $P_{\mathrm{s}}$ pop.  \\    
\hline                        
mean $v_{\mathrm{pec, BJ}}$ ($ \mathrm{km}\, \mathrm{s}^{-1}$)    &  $29 \pm 11$ & $16 \pm 8$ \\      
mean $v_{\mathrm{pec, iso}}$  ($ \mathrm{km}\, \mathrm{s}^{-1}$) & $28 \pm 11$ & $16 \pm 8$  \\ 
\hline
\end{tabular}
\begin{tabular}{l l}
\centering
$p$ & 0.004      \\ 
$p_{iso}$  & 0.007    \\ 
pop. $v_{\mathrm{pec, BJ}}$ difference & (4.4, 22.0) \\ 
pop. $v_{\mathrm{pec, iso}}$ difference & (3.5, 20.9) \\ 
$d_{\mathrm{C}}$ & 1.43 \\ 
$d_{\mathrm{C, iso}}$ & 1.32  \\ 
\hline
\end{tabular}
\end{table}

Here and throughout this paper, the quoted $p$-values have the usual statistical meaning: they represent the probability of obtaining a test statistic at 
least as extreme as the observed one when the null hypothesis is correct. In the case of the peculiar velocities of Galactic BeXRBs, the null hypothesis is that the data are drawn from
the same underlying distribution. We adhered to the classical threshold of $p<0.05$ for a significant result.

The divide of $P_{\mathrm{s, split}}$ =~40~s adopted here to split the two subpopulations was estimated from the histogram of $\log \, P_{\mathrm{s}}$ values of the BeXRB pulsar sample studied by KCP. The dip in the $\log \, P_{\mathrm{s}}$ histogram in Fig. 1. of KCP is rather wide, ranging from approximately $P_{\mathrm{s}}$ = 20 to 80~s. To test the robustness of the above results, we repeated the above analysis for the $P_{\mathrm{s, split}}$ of 20, 60, and 80~s using the $v_{\mathrm{pec, BJ}}$ velocities. The recomputed p-values from the Anderson-Darling test, credibility intervals of the velocity differences between the subpopulations, and Cohen's d for each $P_{\mathrm{s, split}}$ are listed in Table~\ref{robustness}. While for $P_{\mathrm{s, split}}$ =~20~s and 80~s the velocity difference is still significant, at $P_{\mathrm{s, split}}$ =~60~s the significance disappeared.

\begin{table}
\caption{Statistical properties of the Galactic BeXRB sample for varying values of $P_{\mathrm{s, split}}$.}             
\label{robustness}      
\centering                          
\begin{tabular}{l l l l}        
\hline\hline                 
$P_{\mathrm{s, split}}$ (s)  & 20  & 60 & 80  \\    
\hline 
$n$ short $P_{\mathrm{s}}$ & 6 & 10 & 11 \\
 $n$ long $P_{\mathrm{s}}$ & 21 & 17 & 16 \\                   
$p$ & 0.006 & 0.12 & 0.019  \\
pop. $v_{\mathrm{pec, BJ}}$ difference & (4.2, 23.7) & (-0.6, 16.2) & (3.2 , 18.9) \\
$d_{\mathrm{C}}$ & 1.54 & 0.75 & 1.18 \\
\hline
\end{tabular}
\end{table}

We also tested an alternative approach of estimating the peculiar velocities by determining the local standard of rest through the proper motions of stars near the estimated distance of the studied system. For each BeXRB in the sample, we queried GDR2 for stars within 30 arcmin radius whose parallax was within 1~$\mathrm{\sigma}$ of the system parallax. This ensured that we obtained $\gtrsim$~1000 stars in the vicinity of each BeXRB that satisfied the parallax criterion. Using the GDR2 proper motions of these stars, we established a local standard of rest by computing the mean and dispersion of these values. Normally, these 'field' proper motion values would be subtracted from the proper motion of the binary, and with a distance estimate, this would be used to compute a peculiar velocity estimate, such as in Kochanek et al. (\cite{kochanek_2019}). However, upon inspection of the field parameters, we found that in general, the dispersions of the proper motion obtained from the stars near the studied systems are too high for any velocity estimates to be meaningful. This was verified when we estimated the peculiar velocity of each BeXRB by drawing 40~000 random samples from the proper motion of its surrounding field, the BeXRB distance distribution, and the BeXRB proper motion distribution (which was assumed to be the normal distribution centered on the GDR2 proper motion value, with the standard deviation being the proper motion error) while also taking the correlations between the proper motions in RA and Dec into account. This yielded peculiar velocity estimates that also supported the hypothesis that there are two kinematic subpopulations of BeXRBs, separated by $P_{\mathrm{s}}$ = 40~s threshold (p $\sim$ 0.01). However, large errors on the parameters of the field population caused these velocities to be overestimated and to be affected by much larger errors than the velocities obtained using the previous method. As a result, these peculiar velocity estimates were discarded and were not considered in the analysis.

\section{Small Magellanic Cloud Be/X-ray binary population}
The SMC contains an unexpectedly high number of BeXRBs. Interestingly, all SMC high-mass X-ray binaries, with the exception of SMC X-1, are BeXRB systems. Therefore and because of its relative proximity, the SMC provides a unique laboratory for studying BeXRBs in a homogeneous and consistent manner (Coe \& Kirk \cite{coe_2015}; Haberl \& Sturm \cite{haberl_2016}).

Still, the SMC distance poses problems for the current astrometric missions. Because of this and its nature as an irregular galaxy, it is not possible to investigate the peculiar velocities of a sufficient number of BeXRBs directly. Using GDR2, Oey et al. (\cite{oey_2018}) studied the kinematics of early-type SMC runaways, including 14 BeXRBs. However,  the $P_{\mathrm{s}}$ for only seven of them is listed in the HMXB catalog of Haberl \& Sturm (\cite{haberl_2016}). It is interesting to note that the mean peculiar tangential velocity of these BeXRBs is $v_{tan, pec} \sim $~30~$ \mathrm{km}\, \mathrm{s}^{-1}$, which is higher than the observed mean peculiar tangential velocity of Galactic BeXRB ($v_{tan, pec} = 15 \pm 6 \, \mathrm{km}\, \mathrm{s}^{-1}$; van den Heuvel et al. \cite{heuvel_2000}). The same result is obtained regardless of the method that is used, and it remains the same whether all BeXRBs are considered or only the X-ray pulsars. A possible reason for this higher velocity may be that the metalicity of the SMC is lower than that of the Milky Way (Renzo et al. \cite{renzo_2019}). Unfortunately, only one BeXRB with $P_{\mathrm{s}} <$ 40~s is included in their analysis. While this system has the lowest peculiar residual velocity of all the sources they studied (not just the BeXRBs), its peculiar velocity as determined locally (using the kinematics of the nearby OB stars) is substantial. Nevertheless, using only one system to characterize a subpopulation is not meaningful.

Therefore, the velocities of a large sample of SMC BeXRBs need to be studied indirectly. In this section we investigate the mutual positions of BeXRBs in the SMC and nearby young star clusters where they might have formed and are now running away after acquiring a high peculiar velocity after the supernova explosion within the progenitor binary. This approach was adopted by Coe (\cite{coe_2005}), who computed a mean peculiar peculiar velocity arising from the supernova explosion for the BeXRBs in the SMC. Using 17 BeXRB pulsars, their value $v_{tan, pec} \sim $~16~$ \mathrm{km}\, \mathrm{s}^{-1}$ is in line with the mean tangential peculiar velocity of Galactic BeXRBs (van den Heuvel et al. \cite{heuvel_2000}). Interestingly, the mean age of the clusters associated with the BeXRB pulsars in Coe (\cite{coe_2005}) is rather high, 130~$\pm$~140 Myr (log($t$/yr)~$\sim$~8.1). While no firm conclusions can be drawn from a value with such an uncertainty, at face value, this seems much higher than the main-sequence lifetime of a B0~V star (the system secondary), even when we consider that the star would be rejuvenated (i.e., its evolutionary clock will be reset) by the mass transfer within the binary, which would still only yield a maximum lifetime of $\sim$ 40~Myr. A possible explanation would be that the cluster ages are only poorly determined, which is common when young star clusters are considered (e.g., Netopil et al. \cite{netopil_2015}). This would also account for the high uncertainty of the mean cluster age derived in Coe (\cite{coe_2005}). Another possibility is that the component masses of the SMC BeXRB progenitors may initially be as low as 7--8 $\mathrm{M_{\odot}}$. It is possible for stars in this mass range to explode as ECSNe, especially if they have lower metallicities, as is the case for the stars in the SMC (Podsiadlowski et al. \cite{podsiadlowski_2004}) During the binary progenitor evolution, the initially more massive star can transfer a substantial amount of mass to the secondary through Roche-lobe overflow, increasing its mass by several $\mathrm{M_{\odot}}$. Its subsequent evolution will be very similar to the evolution of an isolated star with higher mass (e.g., Podsiadlowski et al \cite{podsiadlowski_92}; Pfahl et al. \cite{pfahl_2002}). This would then account for the fact that the spectral distribution of SMC BeXRBs is consistent with that of the Milky Way (Reig \cite{reig_2011}). 

\subsection{Clusters from Rafelski \& Zaritsky}

Recently, there has been a sharp increase in the number of SMC BeXRBs (Coe \& Kirk \cite{coe_2015}; Haberl \& Sturm \cite{haberl_2016}). Therefore, it is worthwhile to repeat the analysis done by Coe (\cite{coe_2005}) on a larger sample size and also look for possible BeXRB subpopulations. We used the catalog of high-mass X-ray binaries in the SMC by Haberl \& Sturm (\cite{haberl_2016}), listing 147 BeXRBs, where we selected all pulsating BeXRBs with precisely determined positions for the further analysis. This selection resulted in a list of 56 sources. Similarly to Coe (\cite{coe_2005}), the SMC clusters comes from the list of Rafelski \& Zaritsky (\cite{rafelski_2005}; hereafter RZ clusters). 

We compared the projected positions of BeXRBs and the RZ clusters. The position of every BeXRB was compared to all RZ clusters and the distance to the closest cluster was obtained. After this, we removed all BeXRBs that lay in the regions that are not covered by the RZ clusters catalog (objects with distances to the closest cluster $>$ 25 arcmin). After this cut, we retained 53 BeXRBs for the analysis.

We caution that this method, while being simple, has significant drawbacks. First, it is in general not possible to establish whether the matched BeXRB/cluster pairs are really equidistant, that is to say, to determine the radial distance offset between them. It is also difficult to determine whether the matched cluster is really the birthplace of the BeXRB, especially when two or more clusters have similar separations from the particular BeXRB. Therefore, automatically picking the closest cluster may not necessarily be correct. Filtering out the old clusters that cannot be the birthplaces of the currently observed BeXRBs alleviates the problem somewhat, but it is apparent that the results obtained using this method need to be interpreted with caution.  More precise proper motions from the future \textit{Gaia} data releases may allow us to confirm the relative system-cluster positions with the peculiar velocities, which would help eliminate some spurious pairings.

  \begin{figure}
   \centering
   \includegraphics[width=\hsize]{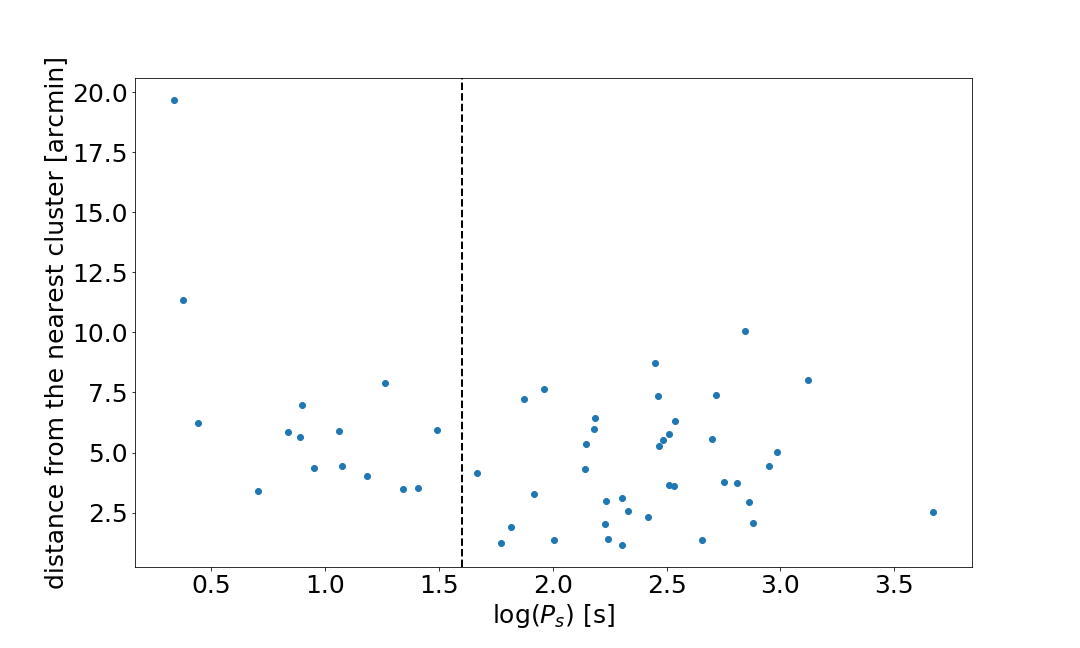}
      \caption{Distances of the SMC BeXRBs from the nearest cluster in the RZ catalog. The conservative estimate of the distance error is 0.7 arcmin, corresponding to twice the median SMC cluster radius listed in the catalog by Glatt et al. \cite{glatt_2010}. Error bars are omitted for clarity.}
         \label{RZ_SMC}
   \end{figure}

Fig.~\ref{RZ_SMC} shows the distances of the studied BeXRBs from the nearest RZ star cluster. The dashed line indicates the approximate boundary between the two subpopulations, where KCP observed a dip in the spin distribution. The subpopulations appear to be marginally kinematically distinct: the short-spin subpopulation have on average larger distances from the star clusters than the long-spin subpopulation. The mean distance from the nearest cluster of the short-spin subpopulation is approximately 6.6$\pm$4.0 arcmin, while for the long-spin subpopulation, it is 4.4$\pm$2.3 arcmin. The reported error values are the standard deviations of the distance distributions. The effect size is significant; the Cohen d is 0.75 between the two subpopulations.

To test the significance of this split in the subpopulations, we conducted a two-sample Anderson-Darling test. The test confirmed that the populations are significantly distinct ($p \, =$~0.039). The bootstrap testing yielded comparable results: the populations are distinct at a credibility better than 95 \%  (but lower than 99 \%).

The values we obtained above are sensitive to the inclusion of the two BeXRBs with the shortest $P_{\mathrm{s}}$, which also exhibit the highest separation from the closest cluster. Excluding them would affect the significance of the subpopulation split, lowering it to $p \, \approx$~0.1. We also used the same cluster catalog as Coe (\cite{coe_2005}), but it can be expected that the results might change if a different cluster catalog were used. Netopil et al. (\cite{netopil_2015}) studied the inferred parameters of Galactic open clusters in different catalogs and found significant dispersion in the ages of open clusters, with the mean standard deviation of approximately 0.5 dex. Even though the open clusters in the SMC are easier to study in some aspects (the distance to the SMC is known, and the reddening is less variable) than the Galactic open clusters, it is likely that there are significant differences between the SMC cluster catalogs. These differences between the star cluster catalogs that are used are likely to have a significant effect on the results of this analysis. We therefore determined the reliability of this result using the more recent SMC star clusters catalogs.

\subsection{Clusters from Nayak et al.}
Nayak et al. (\cite{nayak_2018}) estimated the parameters, including ages, of 174 SMC star clusters. They also collected the parameters of star clusters that were not included in their studied sample, producing a combined catalog of 468 clusters in total, which they used to study the spatio-temporal cluster distribution. We used this catalog to repeat the workflow outlined in the previous section. Because the catalog contains reliable cluster ages, it is possible to study the statistics of clusters/BeXRBs distributions after the age cuts are applied to exclude clusters of a particular age.

To determine the cluster age range in which it is worthwhile to study BeXRB/cluster pairings, we need to estimate the maximum age of the cluster that can be associated with a currently observed BeXRB. This age corresponds to the maximum age of the BeXRB secondary, where it is possible to make a conservative estimate. A star with a mass of 6~$\mathrm{M_{\odot}}$, which is considered to be a low-mass limit for supernovae if the star is in a binary (Podsiadlowski et al. \cite{podsiadlowski_2004}), has a main-sequence lifetime of $\sim$ 110 Myr. If it accretes mass from the primary near the end of its lifetime, it rejuvenates, meaning that it will subsequently evolve like a more massive star, but its evolutionary clock will be reset. As discussed above, we also need to consider the potential uncertainty in estimated cluster ages of approximately 0.5 dex. Thus, in order to include as many viable clusters as possible, it is necessary to include clusters as old as log($t$/yr) = 8.6. Clusters older than this age cut contaminate the analysis.

\begin{figure}
  \resizebox{\hsize}{!}{\includegraphics{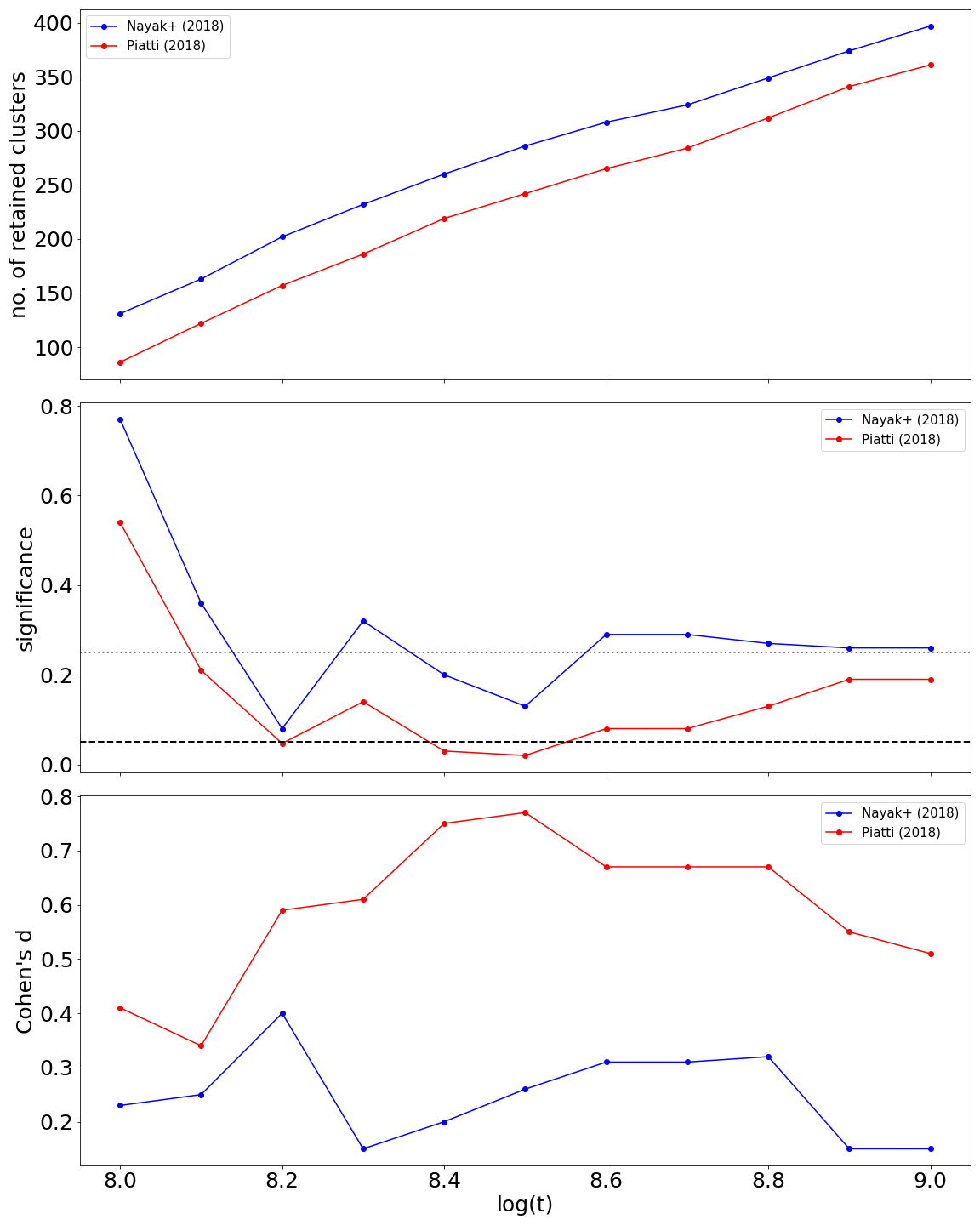}}
  \caption{Top: Number of retained star clusters when clusters older than a particular age limit from the catalogs of Nayak et al. (\cite{nayak_2018}) and Piatti (\cite{piatti_2018}) are excluded. Middle: Significance of the BeXRB--closest cluster distance bimodality as obtained from a two-sample Anderson-Darling test with respect to the applied age cutoff. The horizontal dashed line indicates the classical significance threshold of $p = 0.05$. The $p$-values that exceed 0.25 need to be considered as approximate because they are computed by extrapolation. This threshold is indicated by the horizontal dotted black line. Bottom: Effect size of the bimodality quantified using Cohen's d as a function of the applied age cutoff.}
  \label{cata_comparison}
\end{figure}

We therefore studied cluster age ranges from log($t$/yr) = 8 to 9 because  the number of clusters younger than this is lower than or comparable to the number of the studied BeXRBs and the clusters older than log($t$/yr)~$>$~8.6 are too evolved for a BeXRBs to be associated with them, even after possible age uncertainties are considered. However, in order to investigate the effect of the contamination by the older clusters, the considered age range was expanded somewhat. We conducted a series of statistical tests in which we applied an age cut every time, starting from log($t$/yr) = 8 in 0.1 dex increments and omitted clusters with ages older than this value. As in the previous section, for each BeXRB we searched for the closest cluster within 25 arcmin, collected the distances between them, and compared the BeXRB--closest cluster distance distributions between the spin subpopulations using the $P_{\mathrm{s, split}}=$~40~s as listed in KCP. The number of retained clusters for the analysis, the significance of the difference between the two BeXRB populations resulting from the two-sample Anderson-Darling test, and the effect size estimated using Cohen's d are shown in Fig.~\ref{cata_comparison}.

\subsection{Clusters from Piatti}
Piatti (\cite{piatti_2018}) also compiled a large catalog of SMC star clusters, 411 in total, that includes cluster ages. We repeated the same analysis using this catalog. The results are also included in Fig.~\ref{cata_comparison} to facilitate comparison with the clusters from Nayak et al. (\cite{nayak_2018}).

The catalogs of Nayak et al. (\cite{nayak_2018}) and Piatti (\cite{piatti_2018}) are not completely independent. This is shown in Fig.~\ref{cata_comparison}, where they follow similar trends, which is most clearly visible around log($t$/yr) = 8.3, where the statistics for both catalogs degrade strongly. For both catalogs, the subpopulations appear to be the most distinct overall when only clusters younger than log($t$/yr) = 8.2 to 8.6 are considered, which is reasonable considering the BeXRB lifetimes and likely errors associated with the  cluster ages. When older star clusters are included, the statistics degrade, as is expected when clusters are introduced that are unlikely birthplaces of BeXRBs; they contaminate the sample. 

The evidence for the BeXRB subpopulation difference is stronger in Piatti's \cite{piatti_2018} catalog, where the $p$-value is near the 0.05 level throughout log($t$/yr) = 8.2 to 8.6, and below it at log($t$/yr)~=~8.2, and 8.4 to 8.5. The subpopulation split is less clear using the sample from Nayak et al. \cite{nayak_2018}, but even then, for both of the catalogs and the entire age range considered, we always observed the trend that the short-spin subpopulation has higher separations at face value, even though the difference is not always statistically significant. 

\section{Discussion}
The hypothesis proposed by KCP predicts that the long-spin subpopulation of BeXRBs possesses higher peculiar velocities owing to a higher mass-loss and a stronger kick that is exhibited by a CCSN. On the other hand, if the spin bimodality is caused by different accretion modes, as proposed by Cheng et al. (\cite{cheng_2014}), we are less likely to observe any difference in the kinematics of the BeXRB subpopulations. The observed trends, although marginal, challenge both hypotheses.

Overall, we found that the mean peculiar tangential velocity for the Galactic BeXRBs is $v_{tan, pec} = $~19~$\pm$~11~$ \mathrm{km}\, \mathrm{s}^{-1}$, which is consistent with but somewhat higher than the previous estimates of Chevalier \& Ilovaisky (\cite{chevalier_98}) and van den Heuvel et al. (\cite{heuvel_2000}). We observed that the short-spin subpopulation of the Galactic BeXRBs exhibits higher tangential peculiar velocities than the long-spin subpopulation. This also holds if the divide between the subpopulations $P_{\mathrm{s, split}}$ is shifted within 20 to 80~s, but we caution that if $P_{\mathrm{s, split}}$ =~60~s is adopted, the difference between the populations is not statistically significant. Using the distances of SMC BeXRBs from the nearest star cluster, we observed marginal evidence of the same effect in the SMC, but the size and reliability of the effect depends on the star cluster catalog that is used for the analysis. While some problems affect the method we used to investigate the kinematics of SMC BeXRBs, which makes the result less reliable, it appears possible to conclude that there is some evidence that the BeXRB subpopulations are kinematically distinct when we also consider the result we obtained for
the Galactic BeXRBs,. 

\subsection{BeXRBs as runaway systems}

We considered the origins of the peculiar velocity of a BeXRB. Two mechanisms induce this additional velocity in BeXRBs: the ejection of a high amount of mass by the supernova explosion, and the asymmetries in the explosion that cause the natal kick for the neutron star. The latter of the two is a consequence of the supernova explosion itself and is thought to be highly stochastic and impossible to describe by a simple analytical prescription. However, scaling relations exist that link the neutron star kick velocity to the mass of the neutrino-heated ejecta, explosion energy, and asymmetry, and the progenitor structure (Janka \cite{janka_2017}; Bray \& Eldridge \cite{bray_2016}). The first of the two mechanisms is a consequence of a rapid mass loss from a binary. During the explosion, the primary loses a considerable amount of mass in a very short time, as compared to the orbital period of the binary. Even if the explosion is spherically symmetric with respect to the remnant neutron star, there is a strong asymmetry with respect to the center of mass of the system. This causes the system to recoil, and it can be described analytically (Zwicky \cite{zwicky_57}; Blaauw \cite{blaauw_1961}; Cerda-Duran \& Elias-Rosa \cite{cerda-duran_2018} and references therein).

In a symmetric supernova explosion, the post-supernova binary only remains bound if the system loses less than half of its mass, in other words, when the supernova ejecta $M_{\mathrm{ej}}$ comprises less than a half of the system's pre-supernova mass, $M_{ej}\, < \, (1/2)(M_{\mathrm{1}}+M_{\mathrm{2}})$, where $M_{\mathrm{1}}$ and $M_{\mathrm{2}}$ are the masses of the primary and secondary components, respectively (Blaauw \cite{blaauw_1961}; Boersma \cite{boersma_1961}). When we consider the mass loss at the supernova to be instantaneous, the binary system gains a peculiar velocity, given by 
\begin{equation}
v_{\mathrm{sym}} = \Bigg( \frac{GM_{\mathrm{2}}}{a_{\mathrm{pre-SN}}} \Bigg)^{1/2} \Bigg( \frac{M_{\mathrm{2}}}{M_{\mathrm{1}} + M_{\mathrm{2}}} \Bigg)^{1/2} \Bigg( \frac{M_{\mathrm{1}} - M_{\mathrm{co}}}{M_{\mathrm{2}} + M_{\mathrm{co}}} \Bigg),
\label{symmetric_kick}
\end{equation}
where $M_{\mathrm{co}} = M_{\mathrm{1}} - M_{\mathrm{ej}}$ is the mass of the compact object (a neutron star in the studied BeXRBs), $G$ is the gravitational constant, and $a_{\mathrm{pre-SN}}$ is the pre-supernova semimajor axis of the binary. The last term on the right-hand side of the equation,
\begin{equation}
 \frac{M_{\mathrm{1}} - M_{\mathrm{co}}}{M_{\mathrm{2}} + M_{\mathrm{co}}} = e,
\label{eccentricity}
\end{equation}
 is the eccentricity of the remnant BeXRB (Iben \& Tutukov \cite{iben_97}).

The acquired peculiar velocity (and eccentricity) can be higher when the supernova explosion was asymmetric and the compact object received a kick at birth. Depending on the orientation and magnitude of the kick, the acquired peculiar velocity is
\begin{equation}
v_{\mathrm{asym}} = \bigg(v_{\mathrm{sym}}^{2} - 2v_{\mathrm{sym}}v_{\mathrm{k}}\cos\psi + v_{\mathrm{k}}^{2} \bigg)^{1/2},
\label{asymmetric_kick}
\end{equation}
with
\begin{equation}
v_{\mathrm{k}} = \frac{M_{\mathrm{co}}w}{M_{\mathrm{co}} + M_{\mathrm{2}}}, 
\label{asymmetric_kick_det}
\end{equation}
where $v_{\mathrm{sym}}$ is given by Eq.~\ref{symmetric_kick}, $w$ is the kick velocity of the nascent neutron star, and $\psi $ is the angle between the kick velocity and the pre-supernova relative orbital velocity.  Equation~\ref{asymmetric_kick} shows that the highest peculiar velocity is attained if $\cos \psi$ =~-1, that is, if the direction of the kick at the supernova explosion was opposite to the direction of the relative orbital velocity (Stone \cite{stone_82}; Gvaramadze et al. \cite{gvaramadze_2011}). In this case, Eq.~\ref{asymmetric_kick} can be rewritten simply as
\begin{equation}
v_{\mathrm{max}} =  v_{\mathrm{sym}} + v_{\mathrm{k}} = v_{\mathrm{sym}} + \frac{M_{\mathrm{co}}w}{M_{\mathrm{co}} + M_{\mathrm{2}}} 
\label{max_velocity}
.\end{equation}

Equation~\ref{symmetric_kick} shows that the BeXRB attains the highest peculiar velocity if the binary is as tight as possible before the supernova explosion. Higher relative mass-loss from the supernova explosion will also yield higher peculiar velocities. Another way to increase the peculiar velocity is by the natal kick of the neutron star (however, these kicks can also counteract the kick resulting from a symmetric explosion; Kalogera \cite{kalogera_1998}). The magnitude of this natal kick can potentially be very high (as evidenced by the high velocities of isolated radio pulsars; e.g., Hobbs et al. \cite{hobbs_2005}), but given its random orientation to the pre-supernova relative orbital velocity and the canonical mass of the nascent neutron star of about $M_{\mathrm{co}}$ = 1.4 $\mathrm{M_{\odot}}$, its impact on the attained systemic velocity is reduced because the neutron star has to drag its massive secondary along, provided that the system has remained bound after the supernova. The contribution of the neutron star kick to the systemic velocity abates with the increasing mass of the secondary component as $M_{\mathrm{co}}/(M_{\mathrm{co}}+M_{\mathrm{2}}) <<$~1.

We caution that BeXRBs are fundamentally a biased population and cannot, in general, be used to make inferences about the general properties of neutron stars and supernovae. Renzo et al. \cite{renzo_2019} estimated that about $\sim$86~\% of the binary systems are disrupted at the moment of the first supernova that occurs within the system, primarily because of the high natal kick that is imparted to the nascent compact object. Eldridge et al. (\cite{eldridge_2011}) estimated a similar disruption rate of $\sim$~80~\%. The majority of the surviving binaries comes from relatively tight pre-supernova orbits, which are less vulnerable to disruption. The majority of these sources is expected to evolve through a phase when they might be detectable as X-ray sources. When Renzo et al. (\cite{renzo_2019}) also included ECSNe in the modeling, allowing for smaller natal kicks imparted to the neutron star, the fraction of bound systems increased significantly from $\sim$14~\% to $\sim$35~\%. 

As outlined above, if a binary survives  the supernova, then its velocity relative to the pre-supernova center of mass changes. Here we make a general and approximate brief estimate of the velocity that is expected to be imparted to the system in the case of the CCSN and ECSN. Podsiadlowski et al. (\cite{podsiadlowski_2004}) postulated that the main factor determining the type of the supernova that the system will undergo is the initial orbital period (or equivalently, the binary separation) of the system at the start of its evolution (see their Fig. 2.). Using the pre-supernova parameters of their prototype binaries and Eq.~\ref{symmetric_kick}, we derive an approximate attained peculiar velocity of the surviving binary system for the CCSN scenario of $\sim$~80~$ \mathrm{km}\, \mathrm{s}^{-1}$ and $\lesssim$~10~$ \mathrm{km}\, \mathrm{s}^{-1}$ for the ECSN scenario. In the CCSN scenario, the attained peculiar velocity of the surviving binary system is predominantly due to the mass loss from the binary after a supernova explosion (the 'Blaauw kick'). However, the impact of the SN asymmetry and neutron star kick also significantly supplements the attained peculiar velocity. If the kick velocity $w$ is drawn from a Maxwellian with $\sigma \sim $~265$ \mathrm{km}\, \mathrm{s}^{-1}$ (motivated by the high peculiar velocities of isolated pulsars; e.g., Arzoumanian et al. \cite{arzoumanian_2002}, Hobbs \cite{hobbs_2005}; Verbunt et al. \cite{verbunt_2017}), this would correspond to a mean kick velocity\footnote{the mean of a Maxwellian distribution is given by $\sigma \sqrt{8/\pi}$} of $w \sim$~420~$ \mathrm{km}\, \mathrm{s}^{-1}$. Even after rescaling $w$ to $v_{k}$ using Eq.~\ref{asymmetric_kick_det}, as the impulse of the kick is shared by the entire system (provided that the binary remains bound; van den Heuvel et al. \cite{heuvel_2000}), this velocity is comparable to $v_{sym}$. However, a kick of this magnitude can lead to the disruption of the binary (Renzo et al. \cite{renzo_2019}), therefore the kick magnitude was likely far lower for the surviving BeXRBs, which in turn decreases its likely contribution to the attained peculiar velocity. In the latter case, we considered the effect of the lower natal neutron star kick of $w \sim$ ~50~$ \mathrm{km}\, \mathrm{s}^{-1}$ (a mean velocity resulting from a Maxwellian velocity distribution with $\sigma \sim$~30$ \mathrm{km}\, \mathrm{s}^{-1}$; Pfahl et al. \cite{pfahl_2002}; Renzo et al. \cite{renzo_2019}), which when 'diluted' by the total mass of the system, is also comparable to the velocity imparted by the binary mass loss due to the supernova.

To facilitate the comparison with the measured tangential peculiar velocities, it is necessary to project these velocity estimates onto a plane, assuming that the velocity distribution is isotropic. This can be done by multiplying the velocities by a factor of $\sim \pi/4$ (e.g., van den Heuvel et al. \cite{heuvel_2000}). This resulted in approximate expected tangential peculiar velocities of $\sim$~60~$ \mathrm{km}\, \mathrm{s}^{-1}$ and $\sim$~6~$ \mathrm{km}\, \mathrm{s}^{-1}$ for CCSN and ECSN scenarios, respectively. Furthermore, it is also necessary to correct for the frame of reference because the estimated velocities are in the rest frame of the pre-supernova binaries and the observed velocities are measured using a frame that corotates with the Galactic disk. When we assume that the BeXRB progenitors have formed in OB associations and clusters, we need to account for the velocity dispersion inside these structures, which is typically $\sim$~5~$ \mathrm{km}\, \mathrm{s}^{-1}$ (e.g. Mel'nik \& Dambis \cite{melnik_2017}). The velocity of the pre-supernova binary inside an OB association and the velocity acquired after the supernova explosion are both randomly oriented, therefore it is necessary to convolve the theoretical velocity distribution by the velocity dispersion. This does not have any significant effect except to smear out the theoretical velocity distribution (see Renzo et al. \cite{renzo_2019}) and slightly increase the expected velocity of the systems that underwent an ECSN. Moreover, Reid et al. (\cite{reid_2014}) noted that star-forming regions may be lagging by about $\sim$~5~$ \mathrm{km}\, \mathrm{s}^{-1}$ behind the rotation of the Galactic disk. When the frame of reference tied to the Galactic disk is used, the peculiar velocities derived in this way are therefore slightly overestimated compared to their true values. However, this systematic shift affects all studied Galactic BeXRB in the same way and is also quite small. For the purpose of our analysis, it can be neglected.  

It needs to be noted that these general and approximate analytic estimates depend heavily on the previous binary evolution. The simulations reported by Renzo et al. (\cite{renzo_2019}) showed that the neutron star natal kick $w$ is the dominant factor affecting the attained peculiar velocity, not the effect of the sudden mass loss from a binary at supernova explosion. The average effective natal kick for compact objects in systems that remained bound after the supernova was $w \sim$~66~$ \mathrm{km}\, \mathrm{s}^{-1}$ for their simulation, where the kicks were drawn from the Maxwellian with $\sigma \sim $~265$ \mathrm{km}\, \mathrm{s}^{-1}$. This is far lower than the average $w$ that would be expected, even when supernova fallback is accounted for. The systems experiencing higher compact object natal kicks were more likely to be disrupted.

When we consider the points above, if the CCSNe and ECSNe both occur in the BeXRB progenitors, we expect according to our analytic estimates to observe two subpopulations with tangential peculiar velocities of about $\sim$~60~$ \mathrm{km}\, \mathrm{s}^{-1}$ and $\sim$~10~$ \mathrm{km}\, \mathrm{s}^{-1}$. Brandt \& Podsiadlowski (\cite{brandt_95}) and Eldridge et al. (\cite{eldridge_2011}) also derived a similar value for the CCSN subpopulation, but Renzo et al. (\cite{renzo_2019}) and Kochanek et al. (\cite{kochanek_2019}) both derived significantly lower values with a median of $\sim$~20~$ \mathrm{km}\, \mathrm{s}^{-1}$ ($\sim$~16~$ \mathrm{km}\, \mathrm{s}^{-1}$ if projected). For systems that underwent an ECSN, both Renzo et al. (\cite{renzo_2019}) and Kochanek et al. (\cite{kochanek_2019}) derived very low peculiar velocities.

A reference frame that corotates with the Galactic disk is prone to many uncertainties. It should still be feasible for modern astrometric missions such as \textit{Gaia}, however, to distinguish between these two subpopulations. 

\subsection{Relationship between the peculiar velocity, eccentricity, and orbital period}

\begin{figure}
  \resizebox{\hsize}{!}{\includegraphics{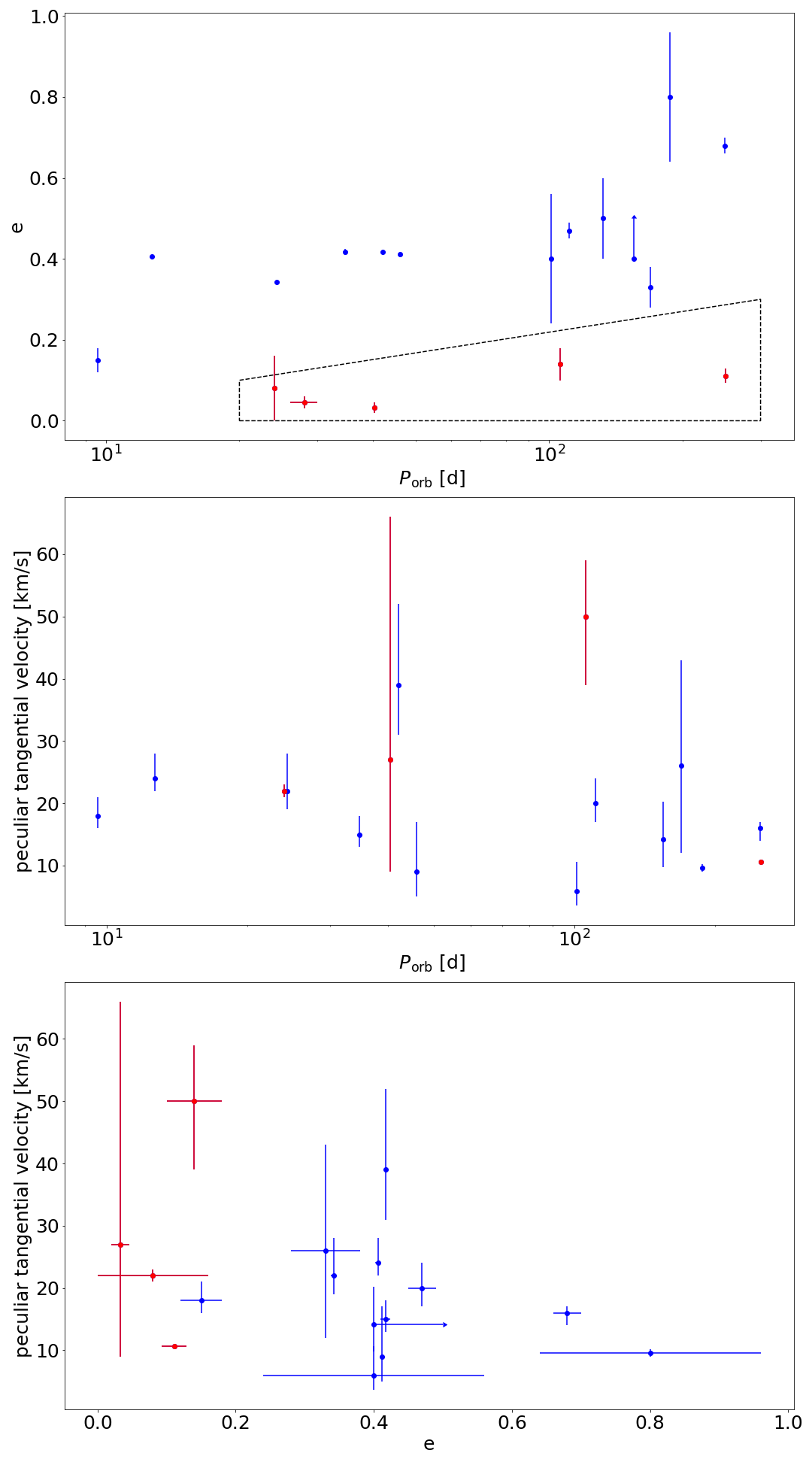}}
  \caption{Relations between the eccentricities $e$, orbital periods $P_{\mathrm{orb}}$ , and peculiar tangential velocities ($v_{\mathrm{pec, BJ}}$) for the sample of Galactic BeXRBs. Top: $P_{\mathrm{orb}}$ vs. $e$ plot for the Galactic BeXRB sample. The region that contains the low-eccentricity class of BeXRBs (Pfahl et al. \cite{pfahl_2002}; the region adopted from Townsend et al. \cite{townsend_2011}) is marked by dashed lines. These systems are marked in red in the current and subsequent graphs for clarity. GX 304-01 and XTE J1906+090 are omitted from the subsequent figure panels because of their unreliable peculiar velocity that is a result of unreliable astrometry. Middle: $P_{\mathrm{orb}}$ vs. tangential peculiar velocity. Bottom: $e$ vs. peculiar tangential velocity.}
  \label{e_vs_speed}
\end{figure}

The orbital solutions available for some of the systems from the Galactic BeXRB population enables us to investigate the relationships between the peculiar velocities, eccentricities and $P_{\mathrm{orb}}$. The top part of Fig.~\ref{e_vs_speed} shows the $P_{\mathrm{orb}}$--eccentricity dependence for the studied BeXRBs, similar to Fig. 6. of Townsend et al. (\cite{townsend_2011}). The plot shows two populations, a low-eccentricity (with $e \lesssim 0.2$) and a high-eccentricity population, which contains most of the remaining systems. Excluding the low-eccentricity subpopulation from the analysis, Townsend et al. (\cite{townsend_2011}) reported a weak correlation between the log($P_{\mathrm{orb}}$) and eccentricity. It is not meaningful for us to perform a detailed correlation analysis because our sample is smaller because we lack the SMC systems. The middle part of Fig.~\ref{e_vs_speed} shows the $P_{\mathrm{orb}}$ and peculiar velocity relationship. These two parameters appear to be uncorrelated or independent of each other. This contradicts the result of the simulations by Brandt \& Podsiadlowski (\cite{brandt_95}), who predicted an anticorrelation between the peculiar velocity and post-supernova $P_{\mathrm{orb}}$.

Recently, Bray \& Eldridge (\cite{bray_2016}) proposed a simple relation for the neutron star natal kick velocity, which is defined as a function of the ratio of the supernova ejecta to the neutron star mass. When the resulting neutron star mass is considered to be about identical for all supernova scenarios, this would mean that the resulting velocity, which is a function of the ejected mass, would in general be  proportional to the neutron star natal kick, which affects the post-supernova eccentricity. Moreover, the scaling relations from Janka (\cite{janka_2017}) suggest the same outcome: a more powerful explosion results in higher peculiar velocities and eccentricities. Therefore, we expect to observe that in general, a high-eccentricity system also exhibits high peculiar velocity. The relation between the orbital eccentricities and peculiar velocities of the systems we studied is shown in the bottom part of Fig.~\ref{e_vs_speed}. The expected relation is clearly not confirmed observationally, and it may even show the opposite trend to what would be expected if the data were taken at face value. This is puzzling, unless some post-supernova orbital circularization mechanism operates in these objects. This is not expected because as outlined above, the timescale for tidal circularization for main-sequence binaries with $P_{\mathrm{orb}}$~$>$~16~d is at least a few tens of Myr, which is significantly longer than the BeXRB phase lifetime (typically $\sim$~10~Myr; van den Heuvel et al. \cite{heuvel_2000}). However, for the systems with $P_{\mathrm{orb}}$~$\lesssim$~16~d, the orbits might be partially circularized, especially if we observe the system near the end of its lifetime. Alternatively, it is possible that the orbit circularizes when the neutron star interacts with the decretion disk of the Be star at periastron passage (e.g., Martin et al. \cite{martin_2009}). Therefore, it is prudent to regard the observed eccentricities as the lower limits of the post-supernova eccentricities. 

However, there is also a high scatter in the data, and possible selection effects also need to be taken into account. Furthermore, we also need to consider (see Eqs.~\ref{symmetric_kick} and ~\ref{eccentricity} and also Maccarone et al. \cite{maccarone_2014}) that the eccentricity induced on a system by a spherically symmetric mass-loss event alone is normally rather low for BeXRBs ($e<0.2$ and far below this value in most cases), while the effect of the neutron star natal kicks is thought to have more impact on the orbital eccentricity, if it is present (van den Heuvel \cite{heuvel_2004}). When the contribution of the neutron star natal kick $w$ to the attained peculiar velocity is significant, it can be expected for a given $w$  that systems with higher eccentricities exhibit lower peculiar velocities and vice versa, as the energy released by the neutron star natal kick can go either into the kinetic energy of the binary system or into the orbit of the system, which would cause the orbit to become larger and/or more eccentric. Therefore, the orbital energy of the binary and its kinetic energy as a whole would be anticorrelated, meaning that the peculiar velocity, eccentricity, and orbital period are linked. This also connects to the previous part of Fig.~\ref{e_vs_speed}, where it seems that the log($P_{\mathrm{orb}}$) is not correlated with peculiar velocity, therefore the orbital energy seems to be increased by pumping the eccentricity rather than widening the binary system.  It is not realistic to expect that all sample systems have received a kick of equal magnitude. Moreover, the quality of the current measurements, the low significance of the correlation, and the number of the systems in the sample are not sufficient to reach any firm conclusions at this point.

While the majority of BeXRBs attained their peculiar velocities at the supernova explosion, as indicated by their young kinematic age compared to the OB runaway stars (Huthoff \& Kaper \cite{huthoff_2002}; Bodaghee et al. \cite{bodaghee_2012}), it is possible that the sample contains some systems that attained some peculiar velocity prior to the supernova explosion, when their progenitors were dynamically ejected from their parent clusters or associations (Poveda et al. \cite{poveda_67}). This two-step ejection scenario (Pflamm-Altenburg \& Kroupa \cite{pflamm_2010}) may be responsible for the higher observed peculiar velocities of some systems, possibly accounting for some outliers.

\subsection{Origins of the kinematic bimodality}

When we consider the points above, there are two possible ways to explain the higher peculiar velocities of the short-spin subpopulation, still within the framework of the different supernovae hypothesis. They might arise from tighter binary progenitors or experience greater relative mass-loss and kicks during the supernova explosion than their counterparts in the long-spin subpopulation. However, this is at odds with the current supernova models, where the ECSNe are not expected to cause high mass-loss and neutron star kicks. 

In the previous sections, we assumed that the short-spin subpopulation arises from the ECSNe and the long-spin subpopulation arises from systems that underwent a CCSN, which is the hypothesis put forward by KCP. This disagrees with the scenario proposed by  Podsiadlowski et al. (\cite{podsiadlowski_2004}), where the short-orbit subpopulation (therefore the short-spin subpopulation because $P_{\mathrm{orb}}$ and $P_{\mathrm{s}}$ are correlated, see, e.g., Fig. 1. in KCP) is instead expected to arise from CCSN. The higher peculiar velocities of the short-spin subpopulation would therefore be explained if this model were adopted instead: the opposite of the hypothesis put forward by KCP. However, we note that the predicted peculiar velocities for the CCSN BeXRB subpopulation, derived from the prototype binaries of Podsiadlowski et al. (\cite{podsiadlowski_2004}; which are simplifications of the reality), seem to be notably higher than what is observed.

The implications are interesting when we assume that the bimodality in the spin period is caused by different accretion modes of the neutron stars in BeXRBs. According to Cheng et al. (\cite{cheng_2014}), the bimodal distribution of the spin period is not directly linked to the two supernova channels, but the supernova mechanism can still be one of the factors that can modulate it, through its influence on the orbital period, eccentricity, and misalignment of the Be star disk, which are some of the factors that then probably affect the occurrence of the giant outbursts. Haberl \& Sturm (\cite{haberl_2016}) used the SMC BeXRB census to investigate which mechanism is responsible for the bimodal spin distribution. They reported higher long-term variability for the short-spin subpopulation, favoring the accretion mode model of Cheng et al. (\cite{cheng_2014}) as the mechanism behind the bimodality. For a BeXRB to show giant outbursts, the right combination of eccentricity and orbital period is apparently needed (Sidoli \& Paizis \cite{sidoli_2018}; Cheng et al. (\cite{cheng_2014}). The relative contribution of other factors is also unclear, such as the activity of the Be star, to the frequency of giant outbursts (e.g., Ziolkowski \cite{ziolkowski_2002}; Reig \cite{reig_2011}). Because the effect of the supernova explosion mechanism is only indirect, it is probably unlikely that any kinematic bimodality is observable at all. If giant outbursts (producing the systems with shorter spin periods) dominate in low-eccentricity systems, which are thought to have an ECSN origin, it is expected that the short-spin subpopulation also has a lower peculiar velocity than the rest of the population, which is not observed. However, this changes when the contribution of the neutron star natal kick to the peculiar velocity is significant because we would expect the systems with low eccentricities to exhibit higher peculiar velocities in general.

\section{Summary and conclusions}
We investigated the kinematics of BeXRB subpopulations that arise from the neutron star spin bimodality in order to test the supernovae hypothesis proposed by KCP and the accretion mode hypothesis of Cheng et al. (\cite{cheng_2014}). We used the GDR2 astrometry to derive the tangential peculiar velocities for the Galactic BeXRBs. In the SMC we used an indirect approach, where we analyzed the distances of the individual BeXRBs from the nearest star cluster as a proxy for the tangential peculiar velocities.

KCP predicted that the short-spin subpopulation, which arises from ECSNe, has systematically lower peculiar velocities than the long-spin subpopulation, which is produced by CCSNe. However, we found some evidence that the subpopulations are kinematically distinct, but in the opposite way as predicted by KCP. The kinematics of the BeXRB subpopulations is difficult to explain within the supernova hypothesis framework because all possible explanations for the increased peculiar velocity of the short-spin subpopulation are hard to reconcile with the current understanding of the ECSNe and CCSNe. Alternatively, adopting the scenario proposed by Podsiadlowski et al. (\cite{podsiadlowski_2004}), which reverses the hypothesis of KCP, appears to resolve the discrepancy between the BeXRB kinematics and the supernova mechanisms.

Although the results derived here are statistically significant, they should be regarded with caution because the sample sizes and the methods we used are limited. Clearly, more reliable determinations of the star cluster parameters in the SMC will provide a more solid base for the type of analysis conducted in this paper. Furthermore, the future Gaia data releases will provide an improved astrometry that can further constrain the kinematic properties of Galactic BeXRBs.

\begin{acknowledgements}
I would like to thank the anonymous referee for their helpful comments that led to a significant improvement of the paper.
This work has made use of data from the European Space Agency (ESA)
mission {\it Gaia} (\url{https://www.cosmos.esa.int/gaia}), processed by
the {\it Gaia} Data Processing and Analysis Consortium (DPAC,
\url{https://www.cosmos.esa.int/web/gaia/dpac/consortium}). Funding
for the DPAC has been provided by national institutions, in particular
the institutions participating in the {\it Gaia} Multilateral Agreement. This research made use of the Python language (van Rossum \cite{van_rossum_95}) and the following packages: NumPy (Oliphant \cite{oliphant_2006}); SciPy (Jones et al. \cite{jones_scipy}); Astropy, a community-developed core Python package for Astronomy (Astropy Collaboration et al. \cite{astropy}) and matplotlib (Hunter \cite{hunter_matplotlib}). This research has made use of the SIMBAD database, operated at CDS, Strasbourg, France.
\end{acknowledgements}

%
%

%
%


\end{document}